\documentclass[twocolumn,traditabstract]{aa} % for the long lists of affiliations 

\usepackage[hyperindex,breaklinks=true]{hyperref}
\hypersetup{
  colorlinks   = true, %Colours links instead of ugly boxes
  urlcolor     = blue, %Colour for external hyperlinks
  linkcolor    = blue, %Colour of internal links
  citecolor    = blue  %Colour of citations
}   
\usepackage{natbib,twoopt}                  %% natbib format for A&A and ApJ
\bibpunct{(}{)}{;}{a}{}{,}                  %% to follow the A&A style

\usepackage{graphicx}
\usepackage{tablefootnote}
\usepackage[binary-units]{siunitx}
\usepackage{txfonts}
\usepackage{xcolor}
\usepackage{multirow}
\usepackage{rotating}
\usepackage{orcidlink}
\usepackage{lscape}
\usepackage{subcaption}
\usepackage{placeins}  

\usepackage{amsmath}
\usepackage{amssymb}
\usepackage[T1]{fontenc}
\usepackage[utf8]{inputenc}
\usepackage{bm}
\usepackage{ragged2e}

\begin{document} 

   \title{Planet formation in chemically diverse and evolving discs}

   \subtitle{II. Chemical fingerprints in planetary atmospheres}

   \author{E. Pacetti\inst{1}\fnmsep\thanks{Corresponding author}\and
          D. Turrini\inst{2}\and 
          E. Schisano\inst{1}\and  
          S. Molinari\inst{1}\and
          C. Walsh\inst{4}\and
          C. P. Dullemond\inst{3}\and
          S. Fonte\inst{1}\and 
          R. S. Klessen\inst{3,6}\and
          U. Lebreuilly\inst{5}\and
          P. Hennebelle\inst{5}\and
          S. L. Ivanovski\inst{7}\and
          R. Politi\inst{1}\and
          D. Polychroni\inst{2,8}\and
          P. Simonetti\inst{7,8}\and
          L. Testi\inst{9}\and
          V. Cottini\inst{10}
          }

   \institute{INAF Istituto di Astrofisica e Planetologia Spaziali, Via Fosso del Cavaliere 100, I-00133, Roma, Italy\\
              \email{elenia.pacetti@inaf.it}\and
             INAF Osservatorio Astrofisico di Torino, Via Osservatorio 20, I-10025, Pino Torinese, Italy\and
            Institut für Theoretische Astrophysik, Universität Heidelberg, Zentrum für Astronomie, Albert-Ueberle-Str 2, D-69120 Heidelberg, Germany\and
            School of Physics and Astronomy, University of Leeds, Woodhouse Lane, Leeds, LS2 9JT, UK\and
            Université Paris-Saclay, Université Paris Cité, CEA, CNRS, AIM, 91191, Gif-sur-Yvette, France\and
            Interdisziplinäres Zentrum für Wissenschaftliches Rechnen, Universität Heidelberg, Im Neuenheimer Feld 225, D-69120 Heidelberg, Germany\and
            INAF – Osservatorio Astronomico di Trieste, via G. B. Tiepolo 11, I-34143, Trieste, Italy\and
            ICSC National Centre for HPC, Big Data and Quantum Computing, Via Magnanelli 2, Casalecchio di Reno, 40033, Italy\and
            Dipartimento di Fisica e Astronomia – Università di Bologna, Via Gobetti 93/2, I-40129, Bologna, Italy\and
            Agenzia Spaziale Italiana, Rome, Italy}

    \abstract{Giant planets form in protoplanetary discs, where the coupled dynamical and chemical evolution of gas and solids determines the composition of the material they accrete. We investigate how planet formation and migration in evolving discs shape the primordial elemental makeup of giant-planet atmospheres. Our aim is to establish a multi-element framework linking atmospheric compositions to planets’ formation pathways and the time-dependent chemical properties of their natal discs. We couple one-dimensional models of viscously evolving discs – incorporating radial dust drift and a comprehensive treatment of volatile chemistry – with $N$-body simulations of planetesimals interacting with a growing and migrating giant planet. Four chemical scenarios (inheritance and reset under low and high ionisation) and three representative grain sizes (0.1, 20, and 100~\si{\micro\metre}) are explored. The coupled outputs are post-processed to track the accretion of carbon, oxygen, nitrogen, and sulphur, and to derive atmospheric elemental ratios normalised to stellar values ($\ast$ denotes stellar normalisation). We identify three atmospheric classes corresponding to distinct accretion regimes: gas-dominated (GD), characterised by N/O*~>~C/O*~>~C/N* and unconstrained or substellar S/N* (near-stellar C/S*); planetesimal-dominated (PD), showing N/O*~<~C/O*~<~C/N*, S/N*~$\geq$~C/N*, and C/S*~$\leq$~C/O*; and drift-enhanced (DE), exhibiting N/O*~<~C/O*~<~C/N* and markedly superstellar volatile-to-refractory ratios. N/O*, C/N*, and S/N* vary systematically with migration extent, although degeneracies arise for planets forming beyond the CO and N$_2$ snowlines; C/O* remains largely insensitive. Metallicity alone does not uniquely trace the balance between solid and gas accretion in drift-dominated regimes. Variations in the disc’s chemical state and dust size imprint distinctive volatile-ratio patterns across the atmospheric classes, providing complementary constraints on these disc properties. This multi-element framework establishes predictive trends to guide the interpretation of atmospheric spectra from current and forthcoming facilities, such as JWST and Ariel, in the context of giant-planet formation.}
 

   \maketitle
%
%#########################################

\section{Introduction}\label{sec:intro}

The rapidly expanding exoplanet census\footnote{\url{https://exoplanetarchive.ipac.caltech.edu}} reveals extraordinary diversity in planetary architectures and bulk properties, particularly among close-in giant planets, suggesting a wide range of formation and evolutionary pathways \citep{Winn2015, Dawson2018}. Advances in spectroscopy have provided direct access to planetary atmospheres, enabling measurements of elemental abundances for both volatile and refractory species, including C, O, N, and S \citep[e.g.,][]{Tsai2023, Feinstein2023, Rustamkulov2023, Bonidie2026}. The upcoming ARIEL mission will extend these measurements systematically across large planetary samples \citep{Tinetti2018}, offering a unique opportunity to use planetary compositions as tracers of planet formation. Yet, linking mature atmospheres to processes in their natal discs remains a major challenge \citep[e.g.,][]{Molliere2022, Feinstein2025}.

Elemental abundance ratios in planetary atmospheres have long been proposed as diagnostics of planet formation, based on the idea that planets inherit the composition of the disc at the location where they accrete most of their mass \citep{Madhusudhan2011a}. The carbon-to-oxygen ratio (C/O) was the first widely adopted tracer of formation location, motivated by the condensation of key volatile carriers (H$_2$O, CO, and CO$_2$) at distinct snowlines, which produces systematic radial variations in the composition of disc gas and solids \citep{Oberg2011}. However, single ratios are insufficient to uniquely constrain formation histories, which involve complex, dynamic, and interdependent processes \citep[e.g.,][]{Mordasini2016, Cridland2019a}. This has motivated multi-element approaches, combining volatile and refractory species to provide more robust constraints on planetary origins \citep[e.g.,][]{Oberg2019, Cridland2017b, Cridland2020, Schneider2021b, Turrini2021, Bitsch2022, Pacetti2022, Chachan2023, Crossfield2023, Nakazawa2026}.

In \citet{Turrini2021} and \citet{Pacetti2022}, we investigated giant planet formation via gas and planetesimal accretion in chemically diverse discs, exploring scenarios of molecular inheritance from the prestellar cloud and complete chemical reset within the disc, which lead to different partitioning of elements into molecular carriers \citep{Eistrup2016}. These studies showed that combinations of C-, N-, O-, and S-based elemental ratios provide diagnostics of accretion and migration histories that remain robust across disc chemical scenarios. However, these models assumed stationary disc structures and did not account for the effects of mass transport, such as the release of volatiles at snowlines driven by the inward drift of icy pebbles \citep[e.g.,][]{Cuzzi2004, Oberg2016, Booth2017, Booth2019, Mah2023}. Pure pebble accretion models \citep[e.g.][]{Bitsch2015, Schneider2021a, Schneider2021b} have shown that the accretion of such volatile-enriched gas by giant planets provides an atmospheric enrichment channel independent of planetesimal accretion \citep{Pollack1996}. These results demonstrate that disc evolution can directly impact the composition of planetary atmospheres, highlighting the need for models that combine disc chemistry, disc dynamics, and hybrid accretion of gas and solids \citep[e.g.,][]{Danti2023}.

To quantify the chemical impact of disc evolution, in \citet[][hereafter Paper~I]{Pacetti2025} we introduced a time-dependent disc model that couples chemical kinetics with gas and dust transport. We examined scenarios of chemical inheritance and reset, and investigated the role of sub-mm-sized grains in transporting volatiles to the inner disc. We found that even the drift of grains smaller than classical pebbles can drive substantial volatile enrichment of the inner disc gas phase.

In this work, we combine the disc model presented in Paper I with $N$-body planet formation simulations from \citet{Turrini2021} to investigate how disc evolution influences the enrichment of planetary atmospheres and the resulting chemical fingerprints. We consider planets accreting both gas and planetesimals, capturing hybrid accretion scenarios in which planetesimal accretion occurs alongside the accretion of volatile-enriched gas produced by drift-driven ice sublimation. This provides a complementary framework to models primarily focused on pebble accretion \citep[e.g.,][]{Lambrechts2012, Bitsch2015, Schneider2021a}, which can also include accretion of volatile-enriched gas \citep[e.g.,][]{Schneider2021a}, or planetesimal accretion \citep[e.g.,][]{Emsenhuber2021, Turrini2021, Pacetti2022}. We analyse elemental abundance ratios in planetary atmospheres, including C, O, N, and S, and establish a predictive framework for interpreting atmospheric compositions in terms of the underlying disc environment and formation history.

\section{Methods}\label{sec:methods}

This work combines two complementary sets of numerical simulations – disc evolution and planet formation – to investigate how the evolution of the natal disc shapes the primordial composition of giant planet atmospheres. The coupled physical and chemical evolution of the disc was modelled using the time-dependent simulations presented in Paper I, which track the radial transport, diffusion, and chemistry of volatile species in both the gas phase and on the surface of dust grains. The planet formation simulations, taken from \citet{Turrini2021}, describe the growth and inward migration of a Jupiter-mass planet interacting with the disc gas and a population of planetesimals, which are introduced parametrically within the disc model and evolved dynamically via $N$-body interactions. 

Both models and simulation campaigns are extensively described in their respective reference works, and are summarised in Sects.~\ref{sec:discmodel} and \ref{sec:pfmodel}. The coupling of the two datasets was performed with the Python post-processing code {\sc HEPHAESTUS}, which combines the time-resolved chemical abundance profiles of key elements in the disc with the evolving accretion history of the planet to reconstruct the elemental composition of primordial atmospheres (Sect.~\ref{sec:couplingmodel}).

\subsection{Disc model}\label{sec:discmodel}

We numerically integrated the disc physics and chemistry with the {\sc JADE} code (Joint Astrochemistry and Disc Evolution; Paper~I), which uses an operator-splitting scheme to evolve them alternately. The disc structure and evolution were modelled with the {\sc DISKLAB} package\footnote{Access to this private package is available upon request to C.P. Dullemond: dullemond@uni-heidelberg.de
}, while chemistry was evolved using the two-phase chemical kinetics code of \citet{Walsh2015}.

The disc is modelled as an axisymmetric, geometrically thin, non-self-gravitating structure accreting onto a solar-type star. The initial gas surface density profile follows the viscous disc solution of \citet{Lynden-Bell1974}. We adopted a characteristic radius $R_{\rm c}=165$~au and a tapering exponent $\gamma=0.8$ to reproduce the observed surface-density profile of the HD~163296 disc \citep{Isella2016}. The normalisation constant was set to $\Sigma_{\rm c}=3.3835$~\si{\gram\per\square\centi\meter} to yield a total disc mass of $0.054\,M_\odot$, consistent with a Minimum Mass Solar Nebula of comparable extent \citep{Hayashi1981, Turrini2021, Sirono2025}. The dust surface density profile is initialised assuming a global dust-to-gas ratio of 0.01 \citep{Bohlin1978}. This ratio refers only to the initial mass of bare dust grains; as the chemistry evolves, the total dust surface density is updated self-consistently to include the mass contribution of volatile species condensed in the ice phase. The midplane temperature is computed from stellar irradiation and viscous heating \citep{Nakamoto1994, Chiang1997, Dullemond2001}, assuming a constant solar luminosity and a lower limit of 10~K.

The disc evolves viscously as an $\alpha$-disc \citep{Shakura1973}, with a constant $\alpha=10^{-3}$. Gas and dust evolution in the midplane are obtained by numerically solving the relevant transport equations: gas follows the 1D diffusion equation of \citet{Lynden-Bell1974}, while dust evolution accounts for radial drift, advection, and turbulent diffusion as in \citet{Birnstiel2010}. Dust grains are assumed to be spherical, with a bulk density of 2.5~\si{\gram\per\cubic\centi\meter}. We performed three independent sets of simulations, each adopting a single representative grain size of 0.1, 20, or 100~\si{\micro\meter}, carrying the full dust mass associated with the initial dust-to-gas mass ratio. This approach allows us to explore the chemical and dynamical effects associated with different dust-gas coupling regimes. The adopted grain sizes bracket the range expected for early disc phases \citep[e.g.,][]{Bate2022} and correspond to progressively weaker dust-gas coupling and more efficient radial drift. Although recent studies suggest that turbulence can drive grain growth to even larger sizes \citep[e.g.,][]{Marchand2023, Lebreuilly2023, Vorobyov2023, Vallucci-Goy2024}, such conditions would likely amplify the drift-driven enrichment effects already captured by our 100~\si{\micro\meter} scenario.

The disc chemical composition is computed self-consistently with the evolving gas and dust. Total elemental abundances follow the protosolar mixture of \citet{Asplund2021}, corrected for atomic diffusion, and are partitioned into rocks, refractory organic carbon ($C_{\rm ref}$), and volatiles (gas and ice). In this partitioning, rocks are calibrated against solar-system meteorites \citep{Lodders2010, Palme2014}, resulting in the entire sulphur budget, approximately $50\%$ of O, $\sim8\%$ of C, and $\sim3\%$ of N residing in refractory components. $C_{\rm ref}$ initially follows a radial profile reproducing the inner Solar System carbon-erosion trends of \citet{Mordasini2016} and \citet{Cridland2019a}, yielding an additional semi-refractory carbon fraction of $\sim60\%$, consistent with cometary measurements \citep[e.g.,][]{Bardyn2017}. The remaining fractions are initialised in volatile form in the gas phase, with the volatile elemental budget partitioned into molecules according to the molecular ratios of the inheritance and reset abundance sets of \citet{Eistrup2016}, corresponding to molecular inheritance from the prestellar phase or complete chemical dissociation during early disc formation. Each chemical scenario is combined with two ionisation prescriptions: a low-ionisation case driven by short-lived radionuclides (SLRs), and a high-ionisation case including both SLRs and cosmic rays (CRs), yielding four chemical setups \citep{Eistrup2016}.

The midplane chemistry is modelled using a network of 668 species and 8385 reactions, including gas-phase, grain-surface, and gas-grain processes \citep{Walsh2015}. The coupled chemical-dynamical evolution is computed over 3~Myr with a time step of $10^3$~yr. The resulting evolution of the gas-phase C, O, and N abundances is shown in App.~\ref{app:A} for the three simulated grain sizes in the high-ionisation inheritance and reset scenarios. 

The formation of planetesimals is not modelled directly within the disc model (e.g. via streaming-instability models such as \citealp{Drazkowska2016}). Instead, planetesimals are introduced parametrically at a conversion time of $t_{\mathrm{conv}}=10^5$~yr, consistent with timescales suggested by meteoritic constraints \citep{Lichtenberg2023, Sirono2025}. At this time, which marks the onset of planet formation in the $N$-body simulations (Sect.~\ref{sec:pfmodel}), $50\%$ of the mass of condensed material (including icy mantles) is converted into a chemically inert planetesimal reservoir throughout the disc, while the remaining $50\%$ continues to evolve as dust in the disc. The radial surface-density profile of planetesimals is therefore derived from the mass distribution of condensed material in the disc at the conversion time. Thus, planetesimals retain the chemical composition of the ice and refractory phases at their formation location, while their subsequent dynamical evolution and accretion by growing planets are handled by the planet formation model.

\subsection{Planet formation model}\label{sec:pfmodel}

We modelled the formation of giant planets using the $N$-body simulations of \citet{Turrini2021}. Here, we summarise the elements relevant for coupling with the evolving-disc model.

All integrations were performed with {\sc Mercury-Ar$\chi$es} \citep{Turrini2019, Turrini2021, Turrini2026}, a high-performance planet-formation $N$-body code based on the {\sc Mercury}~6 hybrid symplectic integrator \citep{Chambers1999}. {\sc Mercury-Ar$\chi$es} models the mass growth, radius evolution, and orbital migration of giant planets within the core-accretion paradigm, accounting for the dynamical interactions between planets and planetesimals, and the combined effects of gas drag and disc gravity. The model and its numerical implementation are described in \citet{Turrini2021} and summarised in Appendix~A of \citet{Pacetti2022} (see also \citealp{Turrini2026} for a detailed and comprehensive description of the code). 

Each simulation tracks a single giant planet growing and migrating in the disc midplane over 3~Myr, from a Mars-mass embryo to a Jupiter-mass planet. Growth occurs through gas and planetesimal accretion. Planetesimals are represented by a swarm of dynamical tracers with semimajor axes randomly sampled between 1 and 150~au, yielding a spatial density of 2000 tracers per au. The tracers represent planetesimals with a fixed physical radius of 50~km, corresponding to the characteristic size predicted by pebble-accretion models \citep{Klahr2016, Johansen2017}. The dynamical interaction of the tracers with the growing planet, under the combined effects of gas drag and disc gravity, is tracked self-consistently by the $N$-body simulations, including scattering and accretion events. To convert the recorded flux of impacting tracers into a mass flux of accreted planetesimals, each tracer is assigned a mass based on the local planetesimal surface density from the disc model, evaluated within a 0.1~au annulus centred on the tracer’s orbital location.

Planet formation proceeds in two stages. During the core-accretion phase (first 2~Myr; \citealp{Bernabo2022, Sirono2025}), the planet grows from $M_0=0.1\,M_\oplus$ to a critical mass $M_c=30\,M_\oplus$, equally divided between the core and the extended envelope. Radius evolution follows the parametric model of \citet{Fortier2013}, based on the hydrodynamic simulations of \citet{Lissauer2009}. During the subsequent runaway gas-accretion phase (final 1~Myr), the planet reaches a final mass $M_F=317.8\,M_\oplus$ ($1\,M_{\rm J}$) and contracts to $R_F=1.15\times10^5$~km ($1.6\,R_{\rm J}$) \citep{Lissauer2009}.

Orbital migration follows the non-isothermal tracks from the population-synthesis studies of \citet{Mordasini2015}, using a piecewise scheme based on \citet{Han2005} and \citet{Walsh2011}. Migration consists of an initial Type~I migration (accounting for $40\%$ of the total radial displacement during core accretion), followed by a faster Type~I phase and a transition to Type~II migration after gap opening. These latter stages produce the remaining $60\%$ of the total migration. Six migration scenarios were simulated, with initial embryo locations at 5, 12, 19, 50, 100, and 130~au, and a final orbital radius of 0.4~au, enabling exploration of diverse accretion and migration pathways.

\subsection{From discs to planetary atmospheres}\label{sec:couplingmodel}

We extracted each planet’s accretion history – the masses of gas and planetesimals accreted as a function of time and position – from the six $N$-body simulations (Sect.~\ref{sec:pfmodel}) and coupled it with the time-dependent outputs of the twelve {\sc JADE} disc-evolution simulations (Sect.~\ref{sec:discmodel}). This provided a time-resolved record of the material accreted during growth and migration, from which we derived the elemental composition of the gaseous envelope, following the method of \citet{Pacetti2022}, here extended to evolving discs with time-dependent radial abundance profiles.

Post-processing was performed with the Python code {\sc HEPHAESTUS}. At each timestep, {\sc HEPHAESTUS} decomposes the accreted gas and solid masses into elemental contributions of H, C, O, N, and S using the local abundances of their dominant molecular carriers and molecular weights computed by {\sc JADE}. These contributions are integrated over the planet's entire accretion history under the assumption of homogeneous mixing within the primordial envelope, yielding bulk primordial atmospheric abundances. These are then normalised to stellar values to obtain the atmospheric C/O*, C/N*, N/O*, S/N*, and C/S* ratios (the asterisk denotes normalisation).

In this work, solid accretion refers exclusively to planetesimals; direct dust accretion is neglected. We find \citep{Pacetti2025} that 100~\si{\micro\meter}-sized grains behave dynamically like pebbles \citep{Lambrechts2012}. We then assume they are efficiently trapped at pressure bumps, preventing significant accretion once the planet reaches the isolation mass. To allow direct comparison across simulations, we also neglected contributions from smaller grains (0.1 and 20~\si{\micro\meter}). Any dust-delivered solids in these scenarios would in any case be limited to $\sim1.5\,M_\oplus$ at most\footnote{If all locally available dust grains are accreted with the gas, a planet accreting $\sim300\,M_\oplus$ of gas would acquire $\sim3\,M_\oplus$ of dust for a dust-to-gas ratio of 0.01 (neglecting the contribution of ices); this reduces to $1.5\,M_\oplus$ if 50$\%$ of the dust forms planetesimals.}, affecting only the planet starting at 5~au, which accretes $\sim1\,M_\oplus$ of planetesimals (Fig.~\ref{fig:metallicity}; Sect.~\ref{sec:metallicity}). Moreover, most gas accretion in this case occurs inside the H$_2$O snowline ($\sim$2~au), where dust carries only rock-forming elements and the mass fraction of condensed material is about half the stellar metallicity \citep{Turrini2021, Turrini2023}. The maximum dust contribution to the atmospheric metallicity would therefore be $\sim0.75\,M_\oplus$ of refractory material, affecting only S and O (Sect.~\ref{sec:discmodel}) and leaving the overall interpretation unchanged.

\section{Results}\label{sec:results}

\begin{figure*}
\centering
\includegraphics[width=0.95\textwidth]{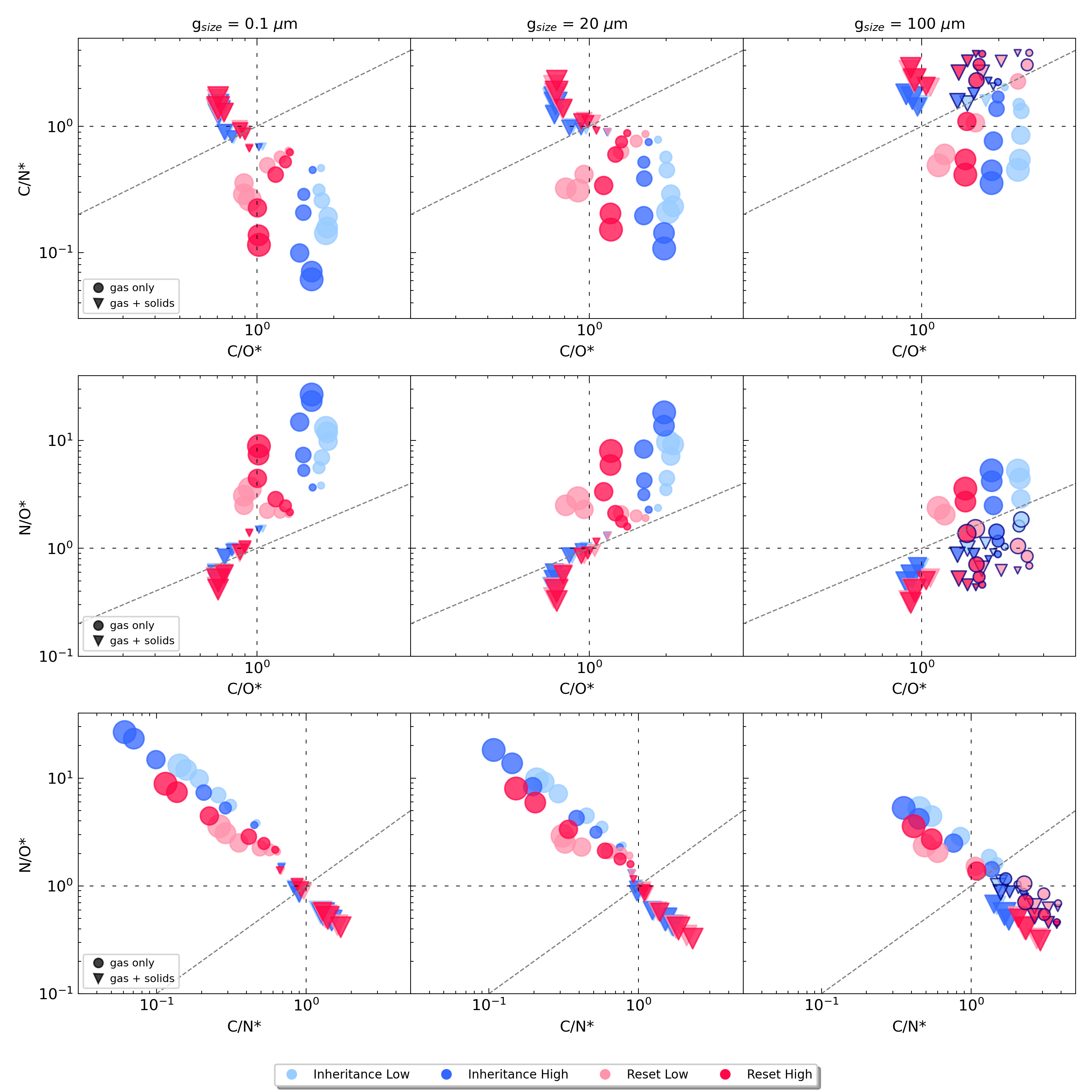}
\caption{Pairwise comparisons of volatile elemental ratios in the primordial atmospheres of the six simulated giant planets. Each panel corresponds to a fixed dust grain size in the natal disc and includes results from all 48 formation scenarios obtained by varying the migration extent (six cases with initial semimajor axes of 5, 12, 19, 50, 100, and 130~au, all ending at 0.4~au), the accretion history of the planet (gas-only or solid-enriched), and the disc’s initial chemical state (four combinations of inheritance vs.~reset chemistry and low vs.~high ionisation). Marker size increases with the total migration distance in each simulation. Circles represent atmospheres dominated by accreted gas, while triangles indicate atmospheres enriched by solids. Points outlined with dark edges correspond to planets formed in the inner regions of discs that are strongly affected by radial drift. Colours denote the disc’s chemical scenario: blue for chemical inheritance (light and dark shades corresponding to low and high ionisation, respectively) and pink for chemical reset (light and dark shades for low and high ionisation, respectively). All ratios are normalised to their respective stellar values; horizontal dashed lines at 1 mark the stellar reference, while the diagonal line in each plot represents the bisector.} 
\label{fig:binary}
\end{figure*}

To investigate and quantify how initial conditions and disc evolution influence the primordial composition of giant-planet atmospheres, we analysed the atmospheric elemental ratios derived as described in Sect.~\ref{sec:couplingmodel}. Figure~\ref{fig:binary} presents pairwise comparisons of the normalised C/O*, C/N*, and N/O* ratios across the four chemical scenarios of the disc (inheritance vs. reset; low vs. high ionisation), three dust grain sizes (0.1, 20, and 100~\si{\micro\meter}), and six initial orbital distances (5, 12, 19, 50, 100, and 130~au). We modelled two classes of planetary atmospheres: ``gas only'', representing planets whose envelope composition is determined exclusively by gas accretion, and ``gas + solids'', representing envelopes compositionally shaped by the accretion of both gas and planetesimals. Results indicate that pairwise comparisons of volatile elemental ratios provide constraints on both disc properties and formation pathways (see below). Hereafter, we refer to the three simulated grain sizes (0.1, 20, and 100~\si{\micro\meter}) as small, intermediate, and large, respectively.

\subsection{Atmospheric signatures of disc properties}\label{sec:res_chem}

Planets characterised by substellar C/N* and superstellar N/O* ratios – all resulting from pure gas accretion (circles in Fig.~\ref{fig:binary}) – are the most sensitive to the disc’s initial chemical state, for which the C/O* ratio provides the strongest diagnostic. In these planets, reset scenarios (in pink) produce systematically lower atmospheric C/O* values than inheritance scenarios (in blue), consistent with the accretion of oxygen-enriched gas typical of reset discs \citep{Pacetti2025}. Quantitatively, pure gas accretion in reset discs produces C/O* ratios ranging from stellar to about $1.5\times$ stellar, whereas it yields superstellar values of $1.5-2.5\times$ stellar in inheritance discs, largely independent of dust grain size. Grain size more strongly affects the C/N* and N/O* ratios, as the lower volatility of C- and O-bearing species relative to N makes these ratios particularly sensitive to ice sublimation effects. Smaller grains correspond to lower C/N* and higher N/O* compared to larger grains. Specifically, regions with C/N*~<~0.2 and N/O*~>~10 are populated exclusively by gas-only planets formed in discs with small or intermediate-sized grains. Among all ratios, the C/O* is the most sensitive to the tested ionisation levels, though the differences remain small: higher ionisation tends to produce a slightly higher (lower) C/O* ratio in the reset (inheritance) scenario, but only for large-scale migrations ($>100$~au) in discs with large grains.

Planets with simultaneously superstellar C/N* and substellar C/O* and N/O* ratios – all solid-enriched (triangles in Fig.~\ref{fig:binary}) – exhibit similar, nearly overlapping trends with migration distance, largely independent of the disc’s initial chemical state, ionisation level, or grain size. This degeneracy suggests that their atmospheric composition is governed primarily by the extent of planetary migration rather than by the underlying disc chemistry. Consequently, observed elemental ratios provide limited leverage to distinguish between disc scenarios in these systems.

The first quadrant of the C/N*–C/O* plane (first row), corresponding to superstellar C/N* and C/O* ratios, is populated exclusively by giant planets formed in discs with large grains (last column). The same applies to the region below the bisector with superstellar C/O* in the N/O*–C/O* plane (second row). Among these, reset scenarios yield higher atmospheric C/N* ratios – about $2-4\times$ stellar – than inheritance scenarios, which remain around $2\times$. The separation is less pronounced in the N/O*–C/O* plane, where results from different chemical scenarios overlap near the stellar value. 

These results demonstrate that pairwise comparisons of volatile elemental ratios provide greater diagnostic power than traditional C/O-based analyses \citep[e.g.,][]{Oberg2011}. When all ratios are considered together, variations in the disc’s initial chemical state and ionisation level can be at least partially constrained in giant planets dominated by accreted gas and in those formed in discs with large grains, while they have a more limited influence on solid-enriched giants formed in discs with small- or intermediate-sized grains. The results for small grains broadly agree with the trends reported by \citet{Pacetti2022} for stationary discs (see Sect.~\ref{sec:discussion} for a detailed comparison).

\subsection{Atmospheric signatures of planet formation}\label{sec:res_pf}

\begin{figure*}
\centering
\includegraphics[width=0.95\textwidth]{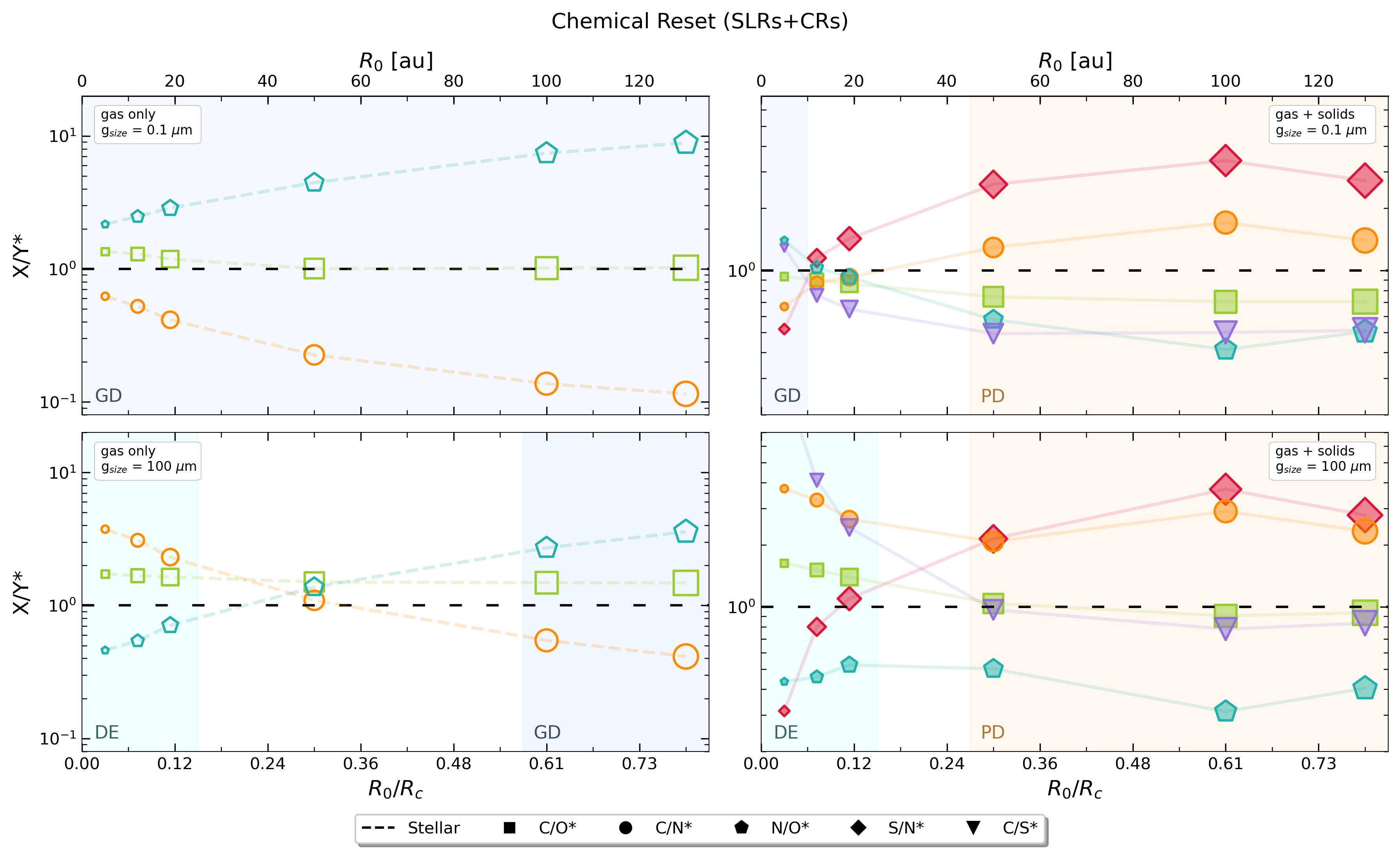}
\caption{Elemental ratios in the primordial atmospheres of the six simulated giant planets as a function of their initial orbital positions. Each planet migrates inward from its initial semimajor axis (5, 12, 19, 50, 100, or 130~au) to a final position at 0.4~au. The x-axis shows the initial semimajor axis, $R_0$, normalised to the characteristic radius of the natal disc $R_{\rm c}$. Marker size increases with the total migration distance in each scenario, while connecting lines are shown for illustrative purposes only. Results correspond to the high-ionisation reset scenario of the natal disc and two representative dust grain sizes: 0.1~\si{\micro\meter} (top row) and 100~\si{\micro\meter} (bottom row). The left panels (open markers) show atmospheres dominated by accreted gas, whereas the right panels (filled markers) show solid-enriched atmospheres. All ratios are normalised to their respective stellar value; the horizontal dashed line at 1 marks the stellar reference in all plots. Coloured shaded regions highlight the characteristic compositional signatures of the three identified classes of giant planets: gas-dominated (GD), planetesimal-dominated (PD), and drift-enhanced (DE). One data point, corresponding to the planet initially at 5~au with a C/S* ratio of $\sim$12, lies beyond the plotted range and is omitted for clarity.} 
\label{fig:ternary}
\end{figure*}

The distribution of atmospheric elemental ratios presented in Sect.~\ref{sec:res_chem} reveals three distinct compositional clusters, each associated with a different dominant channel of heavy-element enrichment during planetary growth and migration – namely, enrichment by unperturbed disc gas, planetesimal accretion, or volatile-enriched gas produced by radial drift – and corresponding to distinct planet-formation pathways. We therefore define three classes of giant-planet atmospheres based on their primary enrichment source: Gas-Dominated (GD), Planetesimal-Dominated (PD), and Drift-Enhanced (DE).

The separation between these classes is illustrated more clearly in Fig.~\ref{fig:ternary}, which combines the elemental ratios shown in Fig.~\ref{fig:binary} with the additional S/N* and C/S* ratios, plotted as a function of the planets’ initial orbital distance. For clarity, the figure presents results for the high-ionisation reset scenario; however, the compositional signatures discussed below apply similarly across the other disc chemical setups (see App.~\ref{app:B}), suggesting that the identified formation signatures are largely insensitive to variations in the disc’s initial chemical state. 

\subsubsection{Gas-dominated atmospheres}\label{sec:GD}
\label{sec:GDplanets}

GD envelopes and atmospheres (blue-shaded in Fig.~\ref{fig:ternary}) correspond to giant planets whose heavy-element budget is acquired predominantly through gas accretion, with absent or negligible contributions from planetesimal accretion and from volatile enrichment produced by ice sublimation from drifting grains. Their atmospheric composition therefore reflects the volatile composition of the disc gas sampled during envelope growth in regions only weakly affected by radial drift. This class corresponds qualitatively to the gas-dominated planets introduced in \citet{Turrini2021} and \citet{Pacetti2022}.

GD atmospheric compositions are characterised by superstellar N/O*, near-stellar C/O*, and substellar C/N* ratios, yielding the characteristic ordering N/O*~>~C/O*~>~C/N*. Within our modelling framework, sulphur resides entirely in refractory phases (i.e., it is delivered to forming planets only through planetesimals); GD atmospheres are therefore also associated with non-detections or strongly substellar values of S-bearing species. The relative enrichment of volatile elemental ratios reflects the intrinsic composition of the disc's volatile reservoir (Sect.~\ref{sec:discmodel}; see also Paper~I). Based on meteoritic data and compositional constraints from the interstellar medium and solar-system comets, our compositional framework adopts a partitioning in which nitrogen is predominantly volatile ($\sim$~97$\%$), while significant fractions of oxygen ($\sim$~50$\%$) and carbon ($\sim$~68$\%$) are sequestered in refractory minerals and semi-refractory organic carbon. Consequently, the disc gas sampled during runaway accretion in regions minimally affected by radial drift beyond the $C_{\rm ref}$ snowline, is intrinsically nitrogen-rich and depleted in oxygen and carbon, resulting in superstellar N/O*, near-stellar C/O*, and substellar C/N* ratios (see App.~C in Paper~I). GD planets therefore inherit this relative enrichment directly from the local disc gas composition at the onset of runaway accretion. These compositional patterns remain robust for discs forming around stars with near-solar compositions, whereas substantially different chemical regimes would naturally require tailored, system-specific modelling.

In our simulations, GD atmospheres arise when runaway gas accretion occurs outside disc regions that are strongly affected by radial drift. This includes gas-only planets forming in discs dominated by small grains (top-left panel in Fig.~\ref{fig:ternary}), as well as planets forming at wide initial orbital distances in discs with large grains (lower-left panel). In the latter case, planets originate beyond the CH$_4$ snowline, at distances greater than $100$~au ($\sim~0.6\,R{\rm c}$) in the scenarios explored, where the gas-phase composition remains largely unaltered by drift-driven volatile enrichment.

The same GD chemical signature is also found in a limited subset of simulations where planets accrete both gas and planetesimals but migrate only short distances ($\lesssim10$~au, or $\lesssim0.06\,R_{\rm c}$) in discs dominated by small grains (top-right panel). In these scenarios, the total mass of accreted solids remains sufficiently small that their contribution does not measurably affect the volatile elemental ratios. These planets display substellar S/N* and near-stellar C/S* ratios, while their C/O*, C/N*, and N/O* ratios follow the same trend as gas-only GD planets.

\subsubsection{Planetesimal-dominated atmospheres}\label{sec:PD}

PD envelopes and atmospheres (orange-shaded in Fig.~\ref{fig:ternary}) form in giant planets that accrete both gas and planetesimals during growth and migration, with their heavy-element budget dominated by planetesimal accretion. In this regime, the refractory and volatile material delivered by planetesimals contributes more than $50\%$ of the total heavy-element content of the envelope, while gas accretion plays a secondary role in setting the envelope metallicity. For the scenarios explored in this work, we find that planetesimals typically contribute at least $70\%$ of the total heavy-element budget (see also Sect.~\ref{sec:metallicity}). These atmospheres are associated with planets that undergo large-scale migration ($\gtrsim 50$~au, or $\gtrsim 0.3\,R_{\rm c}$ in our simulations), enabling prolonged interaction with the disc planetesimal population. This class corresponds qualitatively to the solid-enriched planets identified in \citet{Turrini2021} and \citet{Pacetti2022}.

PD atmospheres are characterised by N/O*~<~C/O*~<~C/N* and a superstellar S/N* ratio. The characteristic relative enrichment of volatile elemental ratios reflects the intrinsic composition of the solid reservoir in the disc (Sect.~\ref{sec:discmodel}; see also Paper~I). Planetesimals formed interior to the N$_2$ snowline are intrinsically nitrogen-poor and enriched in carbon, oxygen, and sulphur relative to the gas. Planets whose heavy-element budget is dominated by planetesimal accretion therefore naturally inherit the relative ordering N/O*~<~C/O*~<~C/N* and elevated S/N* ratios, largely independent of the detailed migration history. 

Most elemental ratios in PD atmospheres deviate from stellar values: S/N* is superstellar, C/N* is near-stellar to mildly superstellar, while C/O*, N/O*, and C/S* are substellar. The resulting S/N*~>~C/N* and C/S*~<~C/O* trends reflect the intrinsically lower volatility of sulphur relative to carbon, oxygen, and nitrogen, such that increasing planetesimal accretion preferentially enhances sulphur relative to the other tracers. In our simulations, this behaviour is most clearly expressed in discs weakly affected by dust drift and ice sublimation, such as those dominated by small grains (top-right panel of Fig.~\ref{fig:ternary}). In these cases, the atmospheric elemental ratios are primarily controlled by the cumulative accretion of solids along the migration pathway.

PD signatures also arise in our disc simulations with large grains (lower-right panel of Fig.~\ref{fig:ternary}), where the sublimation of ice mantles on drifting grains efficiently enriches the inner-disc gas phase in volatiles. Under these conditions, although planetesimals remain the dominant contributor to the metallicity of PD atmospheres, the accretion of volatile-enriched gas modifies the elemental ratios. In the reset scenarios, radial drift preferentially enriches the gas phase in carbon and oxygen relative to nitrogen, because the dominant N-bearing species are more volatile and less efficiently retained in the ice phase across the disc (Fig.~\ref{fig:disc_comp_rh}; see also Paper~I). Compared to the weak-drift regime discussed above, this leads to larger C/N* and lower N/O* ratios in the planetary atmospheres, while S/N* remains largely unaffected, since sulphur resides entirely in the refractory phase in our framework\footnote{In our model, sulphur does not sublimate within the disc regions sampled by the simulations; variations in S/N* therefore arise only from changes in nitrogen abundance.}. These planets display superstellar S/N* and C/N* ratios, typically reaching $\gtrsim 2\times$ stellar for the disc and migration scenarios explored here, near-stellar C/O* and C/S*, and substellar N/O* ratios. Similar trends are found in the inheritance scenarios in our simulations (Figs.~\ref{fig:ternary_il} and \ref{fig:ternary_ih}), although with a different magnitude of enrichment. Inheritance discs retain a larger fraction of nitrogen in the ice phase – primarily as NH$_3$ – which is efficiently transported inward by radial drift (Fig.~\ref{fig:disc_comp_ih}). Consequently, PD atmospheres formed in these discs exhibit comparatively lower C/N* and S/N* ratios than their reset counterparts (typically $\lesssim 2\times$ stellar for the scenarios we explored), while maintaining the same relative enrichment of elemental ratios that defines the PD class.

\subsubsection{Drift-enhanced atmospheres}\label{sec:DE}

DE envelopes and atmospheres (cyan-shaded in Fig.~\ref{fig:ternary}) correspond to giant planets whose atmospheric metallicity is dominated by the accretion of disc gas locally enriched in volatiles by the sublimation of ice mantles on drifting grains at successive snowlines. In this class, atmospheric enrichment arises from volatile-enriched gas rather than from the direct accretion of refractory and volatile material through planetesimals.

In our simulations, DE atmospheres form in discs dominated by large grains and for planets that undergo only limited migration ($\lesssim$~20~au, or $\lesssim$~0.12~$R_{\rm c}$). Under these conditions, planets may also accrete planetesimals, but the short migration distance limits the contribution of solids, leaving volatile-enriched gas as the primary source of atmospheric metallicity.

DE atmospheres are characterised by N/O*~<~C/O*~<~C/N*, similar to the relative enrichment of PD atmospheres, but are distinguished by a markedly lower refractory-to-volatile ratio, as traced by the S/N* ratio or, equivalently, by a markedly higher  volatile-to-refractory ratio, as traced by the C/S* ratio (see also Sect.~\ref{sec:metallicity}). In the gas-only scenario (lower-left panel of Fig.~\ref{fig:ternary}), sulphur-bearing species are not present in the envelope, as planets did not accrete refractory material. When limited planetesimal accretion accompanies gas accretion, these atmospheres exhibit S/N*~<~C/N* and C/S*~>~C/O*, with S/N* decreasing from near-stellar to substellar values as the planet's initial orbital distance decreases, while C/S* becomes increasingly superstellar along the same sequence (lower-right panel). These trends reflect the progressively decreasing contribution of planetesimal accretion for planets originating and migrating from shorter distances. 

As with PD atmospheres, the magnitude of the elemental ratios depends on the disc chemical scenario. In reset discs, gas accreted under efficient radial drift is preferentially enriched in carbon and oxygen relative to nitrogen, because the dominant nitrogen-bearing species remain in the gas phase and are not efficiently released by ice sublimation (Fig.~\ref{fig:disc_comp_rh}). Consequently, DE atmospheres exhibit well-separated elemental ratios with N/O*~<~C/O*~<~C/N*, along with substellar N/O*, near-stellar C/O*, and superstellar C/N* values.

In inheritance discs, a larger fraction of nitrogen is retained in the ice phase as NH$_3$ and is progressively released into the gas at the corresponding snowline (Fig.~\ref{fig:disc_comp_ih}). As a result, DE atmospheres are more nitrogen-rich than in the reset case, leading to a slower decline of N/O* and a more moderate increase of C/N* with decreasing migration distance (Figs.~\ref{fig:ternary_il} and \ref{fig:ternary_ih}). Volatile ratios typically remain between stellar and $2\times$ stellar, approaching the characteristic N/O*~<~ C/O*~<~ C/N* relative enrichment only for the shortest migration scales explored (e.g. planets starting at 5~au). In principle, such nitrogen-enhanced DE atmospheres could be identified observationally through independent constraints on atmospheric nitrogen abundances relative to stellar values.

\subsubsection{Mixed atmospheres}\label{sec:mixed}

\begin{figure*}
\centering
\includegraphics[width=0.9\textwidth]{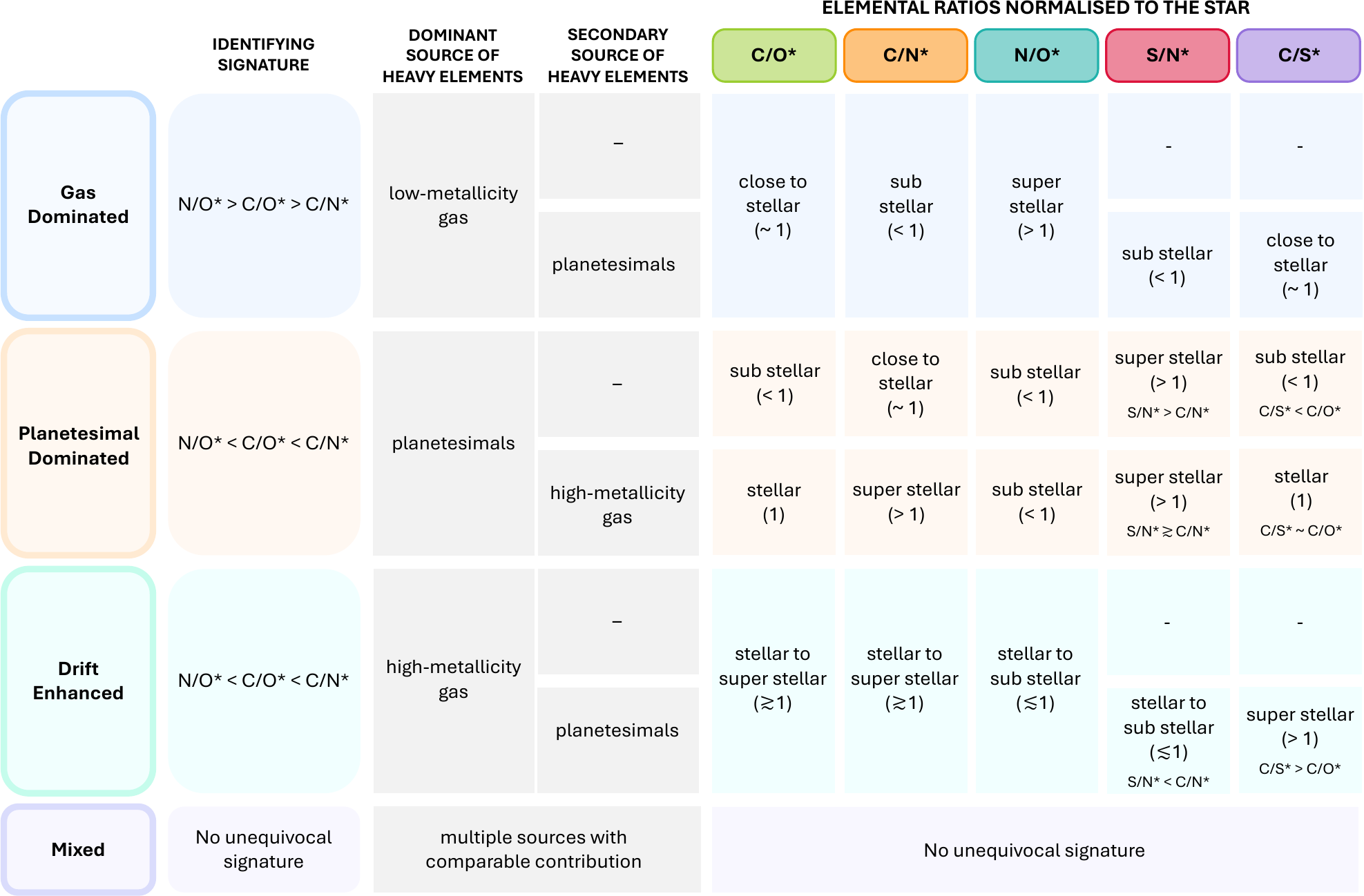}
\caption{Envelope compositions obtained in this study. The table reports, for each atmospheric class, the identifying compositional signature, the dominant source of heavy elements, and a possible secondary source of heavy elements that modifies the elemental ratios without altering the primary enrichment pathway or the identifying signature. The qualitative behaviour of the elemental ratios C/O*, C/N*, N/O*, S/N*, and C/S* is also shown, where the asterisk indicates normalisation to stellar values. Values are reported schematically as relative to the stellar composition or as non-detectable.} 
\label{fig:summary_tab}
\end{figure*}

The three compositional classes described above (GD, PD, and DE) represent end states in which a single source of heavy-element enrichment – unperturbed disc gas, planetesimals, or drift-enhanced gas – dominates the atmospheric metallicity. Between these classes, we identify a population of planets with mixed atmospheres (unshaded region in Fig.~\ref{fig:ternary}), corresponding to transition regimes without a dominant enrichment channel.

Mixed atmospheres are characterised by volatile elemental ratios clustered around the stellar value, with no unequivocal ordering among N/O*, C/O*, and C/N*. This reflects their composite accretion histories, where comparable contributions from multiple sources dilute the distinctive signatures identified for the end-member classes. As a result, mixed atmospheres do not exhibit a unique diagnostic pattern in elemental-ratio space.

In our simulations, mixed atmospheres arise at the boundaries between the GD, PD, and DE regimes. They include planets with envelopes enriched by gas and sublimated ices (gas-only mixed cases; lower-left panel in Fig.~\ref{fig:ternary}), as well as planets that accrete both gas and planetesimals (solid-enriched mixed cases; top-right panel). Gas-only mixed atmospheres are associated with non-detections of S-bearing species, whereas those enriched by solids display near-stellar S/N* and C/S* ratios. In inheritance scenarios (Figs.~\ref{fig:ternary_il} and \ref{fig:ternary_ih}), gas-only mixed atmospheres partially overlap with the compositional space occupied by DE atmospheres, highlighting an intrinsic degeneracy between these regimes when nitrogen-rich ices are efficiently released into the gas phase.

More complex combinations, involving comparable contributions from gas, sublimated ices, and planetesimals to the atmospheric metallicity, although plausible, are not captured by the parameter space considered in this study.
\\

\noindent A schematic summary of our results is shown in Fig.~\ref{fig:summary_tab}. The table presents the envelope compositions of the three atmospheric classes identified in this study (GD, PD, and DE), including their characteristic compositional signature, the dominant source of heavy elements, a possible secondary source of enrichment, and the qualitative behaviour of the normalised elemental ratios. The reported trends reflect the behaviour found across most chemical setups explored in our simulations, while scenario-dependent variations are discussed in the relevant subsections.

We further extended our analysis by considering the elemental abundances C/H*, O/H*, N/H*, and S/H*\footnote{The asterisk denotes normalisation to stellar values.} in the primordial envelope of the simulated planets. We find that the overall compositional picture remains unchanged compared to the elemental-ratio framework discussed above. The same three atmospheric classes (GD, PD, and DE) are identified, and mixed atmospheres remain compositionally degenerate. While elemental abundances provide some complementary information in specific regimes, they do not alter the atmospheric classification scheme or break the main degeneracies identified in the elemental-ratio analysis. A detailed discussion of the atmospheric elemental abundances and their relation to the compositional classes identified in this study is provided in App.~\ref{app:C}.

\subsubsection{Constraints on planetary migration}

The trends in atmospheric elemental ratios also retain information about the extent of planetary migration, although the diagnostic power depends on the atmospheric class and may be affected by degeneracies associated with formation location and enrichment pathway.

In GD atmospheres, N/O* and C/N* progressively deviate from their stellar values with increasing migration distance, while C/O* remains approximately stellar. This behaviour reflects the gradual enhancement of nitrogen relative to carbon and oxygen as planets accrete gas from increasingly volatile-depleted regions of the disc along longer migration pathways. The trend is particularly pronounced in discs with small grains (top-left panel in Fig.~\ref{fig:ternary}), where N/O* rises from about two to ten times the stellar value and C/N* decreases from roughly 0.7 to 0.1 times stellar in our simulations.

A similar monotonic behaviour is observed in PD atmospheres, where N/O*, C/N*, and S/N* increase with migration distance across most scenarios (right-hand panels in Fig.~\ref{fig:ternary}). The trend partially reverses for planets originating beyond the CO and N$_2$ snowlines ($\sim100$~au, or $>0.6\,R{\rm c}$ in our disc), where all major volatile carriers of C, O, and N have condensed into ices and planetesimals retain nearly stellar ratios of C, O, N, and S. Consequently, these planets can exhibit volatile signatures similar to those of PD planets formed further in (e.g., at $\sim50$~au, or $0.3\,R_{\rm c}$), despite exhibiting higher heavy-element enrichments (see Sect.~\ref{sec:metallicity}). 

In DE atmospheres, planets that originate closer to the star exhibit the largest deviations from stellar volatile ratios, reflecting their greater exposure to drift-driven volatile enrichment. As migration distance increases, the elemental ratios progressively converge towards stellar values. In cases with additional planetesimal accretion, S/N* increases from substellar to superstellar values (lower-right panel in Fig.~\ref{fig:ternary}) due to the growing, though still subdominant, contribution of sulphur from planetesimals.

Taken together, these trends show that the combined behaviour of multiple elemental ratios retains information about planetary migration and accretion history. However, the diagnostic power of individual ratios varies across atmospheric classes, and important degeneracies remain for planets forming beyond the major snowlines or experiencing multiple enrichment pathways. This further motivates the use of multiple elemental tracers, rather than individual abundance ratios, when interpreting atmospheric compositions. 

\subsection{Atmospheric signatures with C, O, and S}\label{sec:COS}

\begin{figure*}
\centering
\includegraphics[width=0.95\textwidth]{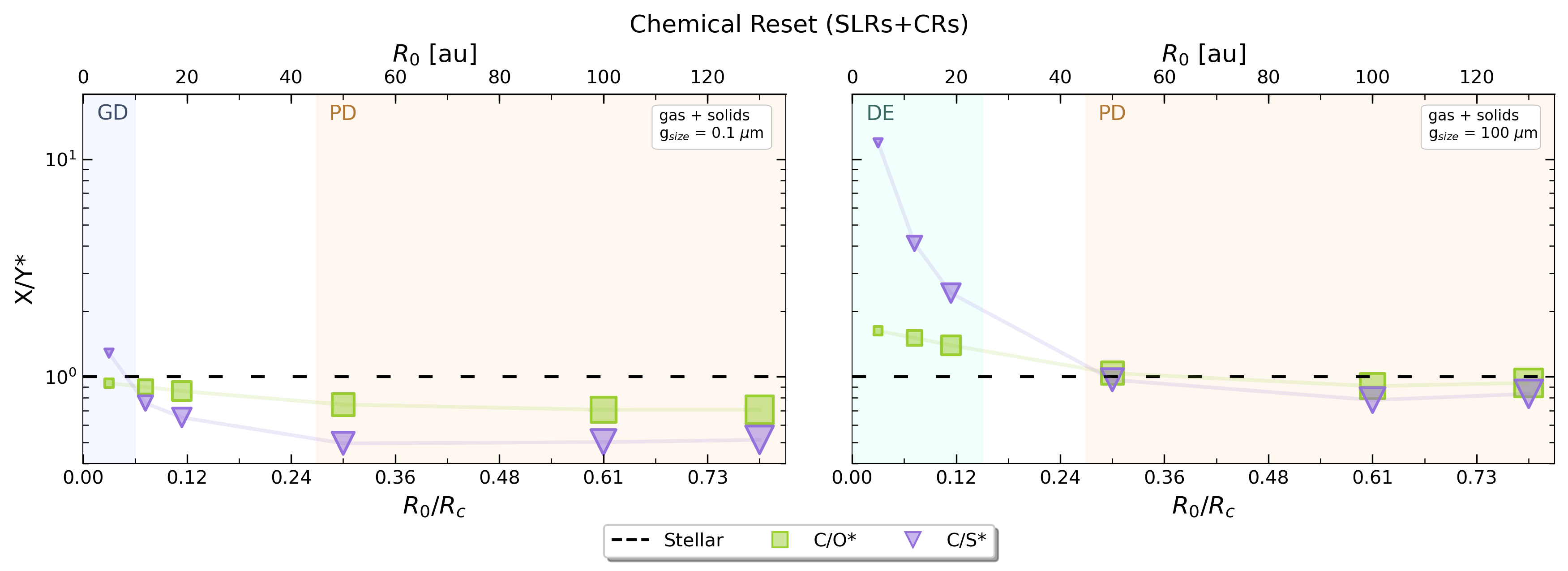}
\caption{Same as Fig.~\ref{fig:ternary}, but showing only the C/O* (green) and C/S* (purple) ratios in solid-enriched atmospheres, for a more direct comparison of elemental ratios when nitrogen-bearing species are not measurable. This subset highlights the characteristic chemical signatures that can still be identified when only C-, O-, and S-bearing species are available for atmospheric characterisation.} 
\label{fig:ternary_COCS}
\end{figure*}

Meaningful fingerprints of planet formation can be identified even when only C-, O-, and S-bearing species are accessible to atmospheric observations, which may often be the case given the limited observability of nitrogen-bearing species \citep[e.g.,][]{Lodders2002, MacDonald2017, Ohno2023}. Figure~\ref{fig:ternary_COCS} illustrates this for atmospheres enriched by both gas and solids, showing that key regions of parameter space remain associated with the main formation pathways even when N-bearing species cannot be constrained.

GD atmospheres and envelopes formed by gas-only accretion are associated with negligible S content (Sect.~\ref{sec:GD}), leaving C/O* as the only accessible elemental ratio when N-bearing species cannot be constrained. The C/O* ratio shows a flat profile as a function of initial semimajor axis (left-hand panels in Fig.~\ref{fig:ternary}), preventing meaningful constraints on migration extent within a C-O-S-only framework. When a minor contribution of planetesimal accretion is included, GD atmospheres (blue-shaded in Fig.~\ref{fig:ternary_COCS}) show C/S*~$\gtrsim$~C/O*, with both ratios remaining close to stellar values. This regime enables the identification of mild solid contamination but provides limited leverage on migration history\footnote{The C/H* and O/H* abundances are more sensitive to migration history than the C/O* ratio. However, their interpretation is complicated by the degree of drift-driven volatile enrichment, which can introduce degeneracies (see App.~\ref{app:C})}.

PD atmospheres and envelopes (orange-shaded in Fig.~\ref{fig:ternary_COCS}) occupy a distinct region of C/O*–C/S* space. In discs weakly affected by radial drift (left panel), where planetesimal accretion dominates the atmospheric enrichment, both C/O* and C/S* are substellar, with C/S*~$\lesssim$~C/O*, reflecting the comparatively larger contribution of S-bearing material from planetesimals. When additional accretion of volatile-enriched gas is present (right panel), both ratios are driven towards stellar values, producing partial overlap with GD atmospheres experiencing modest solid enrichment and indicating that the two pathways may not always be uniquely distinguished without nitrogen-based tracers (see Sect.~\ref{sec:metallicity}).

DE atmospheres and envelopes (cyan-shaded) remain the most readily identifiable class when only C-, O-, and S-bearing species are available. Their atmospheres are characterised by C/S*~>~C/O*, with superstellar C/O* and markedly superstellar C/S* ratios. Importantly, the C/S* ratio remains sensitive to migration history, increasing systematically for planets that originated closer to the star. Our results show that strongly superstellar C/S* ratios, typically associated with accretion of volatile-enriched gas driven by the inward drift of icy pebbles and the release of volatiles at snowlines \citep[e.g.,][]{Schneider2021b, Crossfield2023}, can also arise from the drift of much smaller grains, down to submillimetre sizes (see Sect.~\ref{sec:metallicity}).

Overall, while the unavailability of nitrogen-bearing species reduces the contrast between atmospheric classes and increases the number of degeneracies, broad discrimination between formation pathways remains possible through the combined use of volatile-to-volatile and volatile-to-refractory tracers.

\section{Discussion}\label{sec:discussion}

\subsection{Metallicity and volatile-to-refractory ratios as tracers of planet formation pathways}\label{sec:metallicity}

\begin{figure}
\centering
\includegraphics[width = \columnwidth]{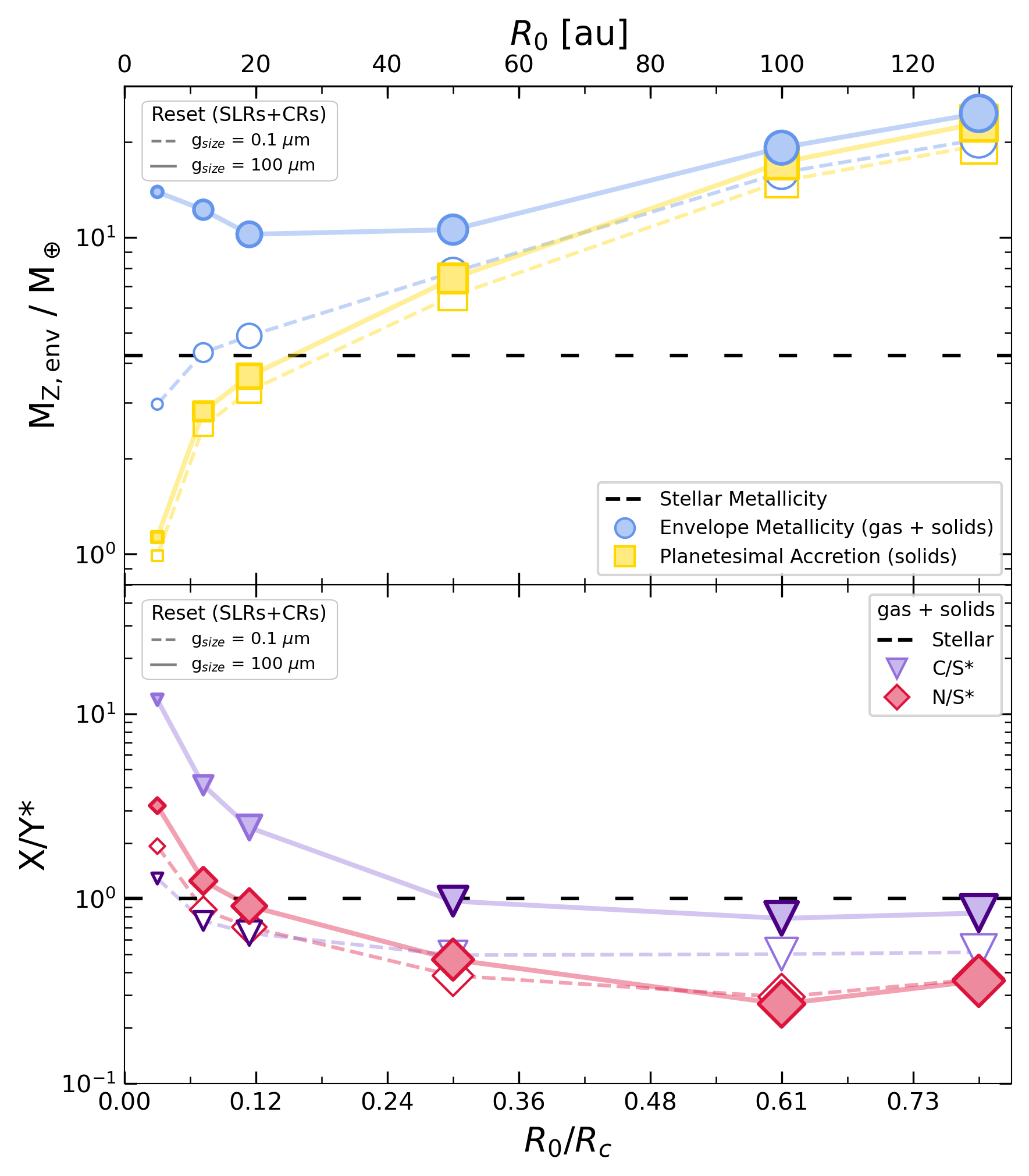}
\caption{\textit{Top}: Mass of heavy elements, defined as all elements heavier than H and He, contained in the primordial envelopes of giant planets accreting both gas and planetesimals and expressed in Earth masses, for the six simulated migration scenarios in the high-ionisation reset scenario of the disc. The shown masses refer only to the envelope composition under the assumption of homogeneous mixing and do not include the additional 15~M$_\oplus$ contained in the planetary core. The x-axis shows the initial semimajor axis, $R_0$, normalised to the characteristic radius of the disc, $R_{\rm c}$. Marker size increases with migration extent, and connecting lines are shown for illustrative purposes only. Blue circles represent the total envelope metallicity (gas + solids), while yellow squares indicate the contribution from accreted planetesimals alone. The horizontal dashed line indicates the heavy-element mass expected for a stellar (solar) metallicity ($Z=0.0141$). \textit{Bottom}: As in Fig.~\ref{fig:ternary_COCS}, but comparing the C/S* with the N/S* ratios. Dark-edged points indicate C/S* ratios compatible with near-stellar values. In both panels, results correspond to the high-ionisation reset scenario and two representative dust grain sizes: 0.1~\si{\micro\meter} (dashed lines, empty markers) and 100~\si{\micro\meter} (solid lines, filled markers).} 
\label{fig:metallicity}
\end{figure}

Previous studies have shown that elemental ratios, particularly volatile-to-refractory ratios, provide a more sensitive diagnostic of the dominant enrichment pathway than the envelope metallicity, allowing one to distinguish whether heavy elements are delivered primarily through volatile-rich gas or solid material \citep[e.g.,][]{Turrini2021, Pacetti2022, Chachan2023, Crossfield2023, Danti2023}. In fact, the envelope metallicity of a giant planet provides an integrated measure of the heavy-element content and does not inform us on which elements dominate the enrichment. For example, the accretion of volatile-rich gas can enrich the envelope primarily through C- and O-bearing species, whereas planetesimal accretion can provide a comparable heavy-element mass through refractory-rich material, including elements such as Fe, Mg, Si, and S \citep[e.g.,][]{Lodders2010}. These pathways can therefore lead to comparable envelope metallicities while producing distinct elemental abundance patterns. Motivated by these results, we investigated how envelope metallicity and volatile-to-refractory ratios behave within the formation scenarios explored in this work.

\subsubsection{Atmospheric metallicity}

The top panel of Fig.~\ref{fig:metallicity} shows how the envelope metallicity of giant planets that accreted both gas and planetesimals builds up in our simulations as a function of initial orbital location. The plot shows the total mass of heavy elements contained in the primordial planetary envelope (blue circles), excluding the core, together with the contribution from accreted planetesimals alone (yellow squares). Both quantities are compared to the mass of heavy elements expected for a solar (stellar) composition, corresponding to an envelope metallicity of $Z=0.0141$ (horizontal dashed line). The figure is the direct analogue of Fig.~6 in \citet{Turrini2021}, which explored similar trends in a static-disc framework.

In discs dominated by submicrometre- and micrometre-sized grains, the envelope metallicity remains a reliable tracer of planetesimal accretion. In this regime (dashed lines, empty markers), the metallicity increases monotonically with migration extent. For planets undergoing large-scale migration ($R_0\gtrsim 50$~au, or $\gtrsim 0.3\,R_{\rm c}$ in our simulations), it closely follows the mass of accreted planetesimals, reflecting their dominant role in enriching the atmosphere. This behaviour is characteristic of the PD scenarios (Sect.~\ref{sec:PD}). For smaller initial orbital distances, the overall trend is preserved, but the two curves progressively diverge. Planets migrating over shorter distances accrete fewer planetesimals, resulting in an increasing contribution of the gas to the heavy-element budget \citep{Turrini2021}. This trend reflects the gradual transition from planetesimal-dominated to gas-dominated enrichment (Sect.~\ref{sec:GD}).

The metallicity trend changes in discs with larger grains (solid lines, filled markers), where volatile-enriched gas provides an additional source of heavy elements to planetary envelopes. For planets undergoing short-scale migration ($R_0\lesssim20$~au, or $\lesssim0.12\,R_{\rm c}$ in our simulations) and for planetesimal contributions up to $\sim4\,M_\oplus$, we find that the envelope metallicity is dominated by volatile-enriched gas and increases as migration distance decreases. This behaviour is characteristic of DE atmospheres (Sect.~\ref{sec:DE}). For larger migration distances and greater accreted planetesimal masses, the solid contribution becomes dominant and metallicity again traces planetesimal accretion, increasing monotonically with migration distance (PD regime)\footnote{The turning point in the metallicity trend depends on the efficiency of drift-driven volatile enrichment, which in turn is influenced by disc properties such as disc radius and mass, viscosity, dust size distribution, and dust-to-gas ratio.}.

As a result, the envelope metallicity is not a monotonic function of either migration distance or accreted solid mass. Planets with very different formation pathways – for example, PD planets that migrated over tens of astronomical units and DE planets that formed close to the star – can exhibit comparable envelope metallicities. This degeneracy confirms that metallicity alone is not a robust tracer of the dominant source of heavy elements once radial transport becomes important, in agreement with the predictions of pebble-accretion models \citep[e.g.,][]{Schneider2021a,Schneider2021b}. Notably, we find that drift-driven volatile enrichment can already produce superstellar atmospheric metallicities and characteristic compositional signatures when transport is dominated by grains as small as $\sim100$~\si{\micro\meter}.

The absolute enrichment levels obtained in our simulations should be interpreted in the context of the adopted disc model. Our reference disc has an initial mass of $0.054\,M_\odot$ and a characteristic radius of 165~au, whereas the pebble-accretion studies of \citet{Schneider2021a} and \citet{Danti2023} adopt substantially more massive and more compact discs, with $M_{\rm d}=0.128\,M_\odot$ and $R_{\rm d}=137$~au, thereby providing a significantly larger solid reservoir and a higher spatial concentration of solids available for inward transport and planetary enrichment. After the conversion of 50$\%$ of the dust into planetesimals, only $\sim85-90\,M_\oplus$ of dust remains available for radial drift in our simulations. Moreover, only the icy component of this reservoir can enrich the gas through sublimation outside the evaporation fronts of refractory species. In our compositional model, ices account for roughly half of the solid reservoir \citep[see also][]{Turrini2021,Turrini2023}, implying that at most $\sim40-45\,M_\oplus$ of heavy elements can be released into the gas phase and subsequently accreted by the planets. The more modest heavy-element enrichments obtained here, compared to those reported in \citet{Schneider2021a} and \citet{Danti2023}, therefore primarily reflect the smaller solid reservoir available in our disc, rather than an intrinsic limitation of drift-driven atmospheric enrichment.

In addition to the available solid reservoir, the atmospheric heavy-element enrichment is also sensitive to the timescale over which drifting solids are supplied to the inner disc \citep[e.g.,][]{Schneider2021a, Mah2023}. In isolated discs, the solid reservoir is finite and the resulting volatile enrichment is therefore a transient phenomenon. As shown in Paper~I (see also Figs.~\ref{fig:disc_comp_ih} and \ref{fig:disc_comp_rh}), the volatile enrichment of the inner disc in simulations with 100~\si{\micro\meter}-sized grains reaches a maximum around 2.5~Myr and subsequently declines as the outer solid reservoir is depleted. Larger particles would accelerate this evolution, shifting the enrichment phase to earlier times and shortening its duration. Consequently, a planet entering runaway gas accretion at a fixed time may accrete either strongly enriched gas or only moderately enriched gas depending on the relative timing between atmospheric accretion and the evolution of the solid flux. The final atmospheric enrichment therefore depends on the interplay between the available solid reservoir, the efficiency of solid transport, the evolution of the disc, and the timing of planetary gas accretion. Comparisons between models adopting different assumptions are therefore not straightforward, and similar drift-enhanced atmospheric signatures do not necessarily imply the same underlying grain-size distribution. 

\subsubsection{Volatile-to-refractory ratios}

A more powerful diagnostic than metallicity is the volatile-to-refractory ratio, which can be traced by ratios such as C/S* or O/S* \citep[e.g.,][]{Turrini2021, Crossfield2023}. In the following, we focus on C/S*, although similar considerations apply to O/S*. The right-hand panel of Fig.~\ref{fig:ternary_COCS} shows that DE atmospheres exhibit superstellar C/S* ratios that are systematically higher than the stellar C/S* ratios characteristic of PD atmospheres. Thus, the C/S* ratio remains a sensitive tracer of whether heavy elements are primarily accreted in solid or gaseous form, even in combined enrichment scenarios.

We note, however, a key difference between our predictions and those of the pebble-accretion models by \citet{Schneider2021b}. Both frameworks predict roughly stellar C/O* ratios and superstellar C/S* ratios for planets accreting high-metallicity gas. However, while \citet{Schneider2021b} predict that C/O* and C/S* increase with initial orbital distance (see also \citealp{Crossfield2023}), our DE atmospheres display a comparatively flat C/O* profile and a C/S* ratio that is highest for planets originating closer to the star (Fig.~\ref{fig:ternary_COCS}). This discrepancy mainly arises from differences in the adopted disc compositional model. Based on the observed carbon depletion in the inner Solar System \citep{Allegre2001, Bergin2015}, our model stores a substantial fraction ($\sim 60 \%$) of carbon in semi-refractory organic material that sublimates over a broader radial range than in \citet{Schneider2021b}, specifically between 1 and 5~au \citep{Mordasini2016}, smoothing radial C/O* gradients in the disc gas-phase (see also Paper~I). At the same time, sulphur remains refractory throughout the disc region sampled by our simulations, so gas accretion closer to the star preferentially enhances C relative to S, leading to elevated C/S* ratios. Both studies therefore demonstrate that volatile enrichment driven by the inward drift of solids and the subsequent release of volatiles at snowlines leaves a clear imprint in giant-planet atmospheres, while also highlighting that well-constrained disc chemical inventories and multiple elemental tracers beyond C/O* alone are required for robust constraints on formation pathways.

As discussed in \citet{Turrini2021, Turrini2022} and \citet{Pacetti2022}, sulphur acts as a proxy for rock-forming elements in the disc – particularly Fe, Mg, Si, and Ni observed in ultra-hot Jupiters \citep[e.g.,][]{Chachan2023, Bonidie2026}. The same qualitative trends analysed in this section are therefore expected for ratios involving other refractory elements, which may in fact provide more direct tracers of rocky material than sulphur, whose chemistry can include minor volatile and semi-volatile reservoirs \citep[e.g.][]{Kama2019,Legal2021,Nakazawa2026}. This implies that the conclusions drawn above extend to a broader class of volatile-to-refractory diagnostics, such as C/Fe or O/Si.

Recent reviews \citep[e.g.,][]{Feinstein2025} have emphasised that volatile-to-refractory ratios constrain the efficiency of drift-driven ice sublimation. We note, however, that not all ratios between a volatile and a refractory element do so with the same sensitivity. The bottom panel of Fig.~\ref{fig:metallicity} compares the behaviour of the C/S* and N/S* ratios in our combined simulations, which include both gas and planetesimal accretion, for discs with small grains (dashed lines, empty markers) and large grains (solid lines, filled markers), corresponding to inefficient and efficient radial drift, respectively. Both ratios increase with decreasing migration distance, reflecting the growing contribution of gas relative to planetesimals to the total metallicity (as shown by the metallicity-planetesimal trend in the top panel). However, their physical interpretation differs substantially.

The C/S* ratio, and similarly the O/S* ratio, responds efficiently to the accretion of volatile-enriched gas produced by the inward drift of icy solids and the subsequent release of volatiles at snowlines, leading to strongly superstellar values. At the same time, near-stellar C/S* values can arise from physically distinct scenarios, including long-range migration with substantial planetesimal accretion in discs with efficient radial transport (dark-edged, filled markers), or more limited migration in discs with inefficient transport and modest planetesimal accretion (dark-edged, empty markers). This introduces a degeneracy in the interpretation of C/S* when considered in isolation.

By contrast, the N/S* ratio is only weakly sensitive to drift-induced volatile enrichment, because N is more volatile than C and therefore less available in the ice phase. For short-scale migrations ($R_0 \lesssim 20$~au, or $\lesssim 0.12,R_{\rm c}$ in our simulations), N/S* follows very similar trends in discs with both inefficient and efficient radial drift. At the same time, even modest planetesimal accretion produces a detectable imprint in N/S*.

As a result, while C/S* is a sensitive tracer of volatile-enriched gas accretion and transport efficiency, N/S* more directly constrains the contribution of planetesimals. Their combined analysis therefore allows us to disentangle the relative roles of gas accretion, radial drift, and solid accretion. This complementarity is particularly important because transport efficiency depends on disc properties such as viscosity, which can alter C/S* trends and mimic variations in accretion history. Anchoring the planetesimal contribution with N/S* thus allows a more robust interpretation of C/S* in terms of drift efficiency, confirming that reliable constraints on formation pathways require the joint analysis of multiple ratios rather than reliance on individual ones.

\subsection{Stationary vs evolving disc}

\begin{figure}
\centering
\includegraphics[width =\columnwidth]{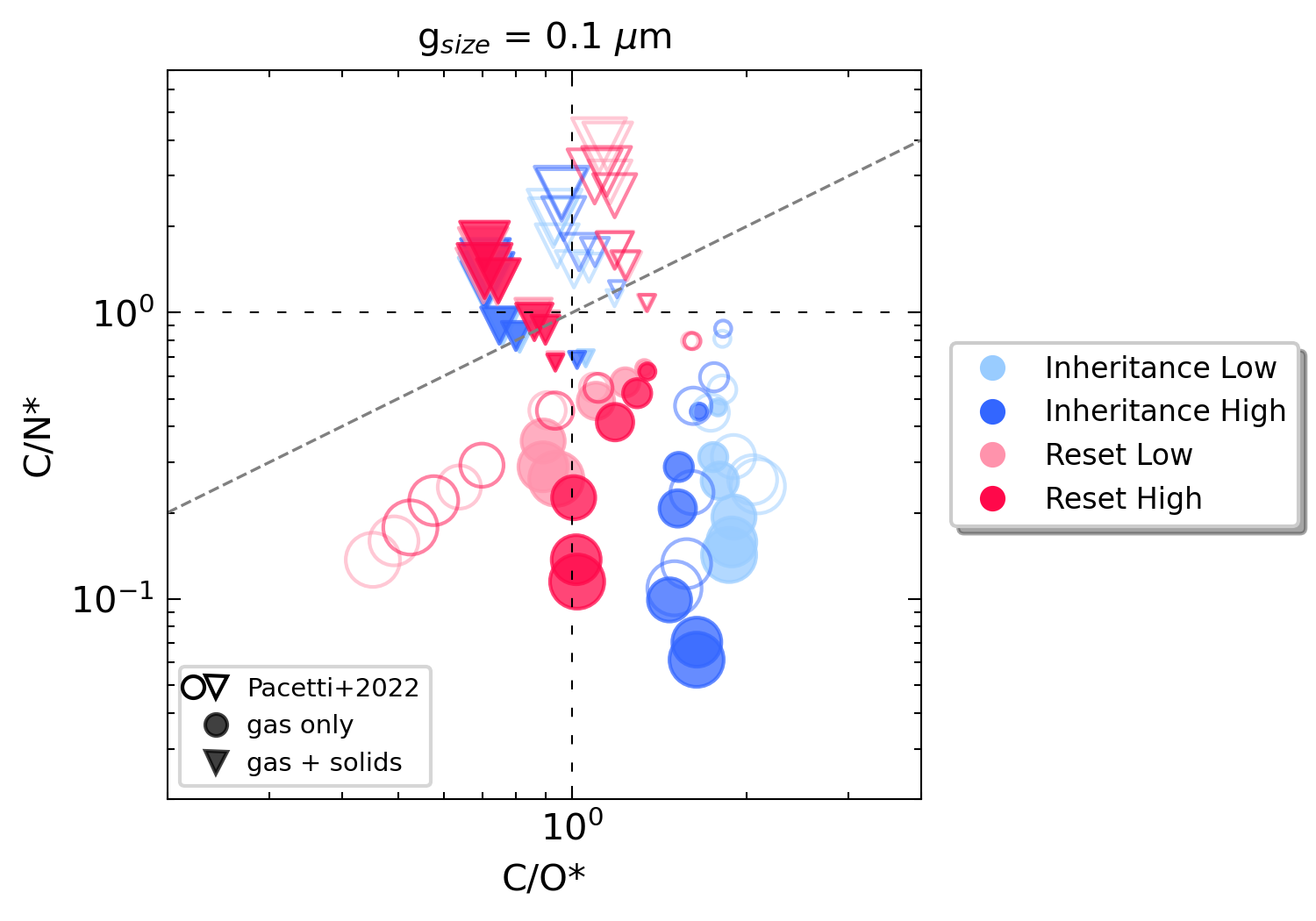}
\caption{Same as Fig.~\ref{fig:binary}, but including the dataset from \citet{Pacetti2022} for stationary discs (open markers).} 
\label{fig:compP22}
\end{figure}

Figure~\ref{fig:compP22} compares the results of this work for discs with 0.1~\si{\micro\meter}-sized grains with those of \citet{Pacetti2022}. Both studies investigate the same disc chemical scenarios. In \citet{Pacetti2022}, the disc was assumed to be stationary, with abundances set by 1~Myr of chemical evolution based on the same chemical network implemented in {\sc JADE}, but without concurrent gas or dust transport \citep{Eistrup2016, Eistrup2018}.

A one-to-one quantitative comparison of atmospheric compositions is not possible because of different physical assumptions in the two models. First, in this work only 50$\%$ of the dust mass is converted into chemically inert planetesimals, whereas \citet{Pacetti2022} assumed complete conversion, effectively doubling the ice inventory available for accretion through planetesimals. Second, here the temperature profile is computed self-consistently from stellar irradiation and viscous heating, while \citet{Pacetti2022} adopted a thermal structure parametrised on the Minimum Mass Solar Nebula \citep[MMSN;][]{Hayashi1981}.

Despite this, clear qualitative trends emerge. For planets that accreted both gas and planetesimals (triangles), the overall C/N*–C/O* distribution is preserved but shifted in absolute values. This offset reflects the different assumptions on the efficiency of dust–planetesimal conversion, indicating that the volatile inventory locked in planetesimals plays a primary role in shaping atmospheric abundances in these scenarios.

For planets whose metallicity is dominated by accreted gas (circles), the comparison depends on the chemical scenario. Under chemical inheritance, the qualitative trends are preserved, with only modest quantitative differences. These arise from gas transport and volatile redistribution in evolving discs where a fraction of the dust mass is removed into planetesimals, progressively depleting the gas phase in volatiles over time \citep[see App.~A in][]{Pacetti2025}. 
By contrast, in reset models, differences in thermal structure and disc evolution lead to both qualitative and quantitative divergences, highlighting the importance of modelling disc structure self-consistently, based on stellar properties, when interpreting formation pathways in reset discs.

Stationary MMSN-like disc models therefore remain a valid first-order approximation under chemical inheritance and limited radial dust transport. However, in chemically reset scenarios, well-constrained disc structures are essential for robust interpretation of planetary atmospheric compositions.

\subsection{Caveats of the model assumptions}

This work aims to investigate how the coupled evolution of disc chemistry, gas and dust transport, planetary growth, and planetary migration shapes the primordial atmospheric composition of close-in giant planets, and to identify the characteristic compositional signatures associated with different formation pathways. However, the complexity of the processes involved makes the inversion from atmospheric abundances to formation pathways intrinsically challenging \citep[e.g.,][]{Molliere2022}. In this study, we focused on isolating the impact of chemical diversity in the disc, different dust-gas coupling regimes, and different accretion and migration histories. A broader exploration of the parameter space will nevertheless be required to assess the robustness of the trends and compositional signatures identified here.

For example, the viscosity parameter $\alpha$ (set to $10^{-3}$ in this study) influences both the chemical evolution of the disc and the growth and migration history of the planets. Lower viscosities reduce the gas accretion rate through the disc and onto the star. In a drift-dominated regime, this would tend to maintain local volatile enrichments near snowlines for longer timescales before they are redistributed by viscous transport, potentially increasing the amount of volatile-enriched gas accreted by planets migrating through these regions \citep[e.g.,][]{Mah2023}. At the same time, lower viscosities slow planetary growth and migration during the runaway gas-accretion phase \citep[e.g.,][]{Danti2023}. This could cause planets to cross fewer evaporation fronts and limit their ability to settle into a final orbit in the innermost disc regions. Qualitatively, we expect the DE atmospheres identified here to be the most sensitive to viscosity variations, whereas the compositional trends identified for GD and PD atmospheres should remain comparatively more robust.

A second caveat concerns the formation of giant-planet cores at large orbital distances. In this study, we assume that planetary embryos form and subsequently grow to a solid core of 15~$M_\oplus$ before entering runaway gas accretion. At the onset of runaway accretion, the planet also contains an additional 15~$M_\oplus$ in the form of an extended gaseous envelope, for a total mass of 30~$M_\oplus$. Whether such cores can form beyond $\sim$10 au within typical disc lifetimes remains an open question. Classical planetesimal-accretion models generally encounter growth-timescale difficulties at large orbital distances. At the same time, ALMA observations reveal rings and gaps over a wide range of orbital distances, indicating that solids can be efficiently concentrated and processed even in the outer regions of protoplanetary discs \citep[e.g.,][]{Bae2023}. Several studies have also suggested that Jupiter itself may have formed in the outer Solar System \citep[e.g.,][]{Pirani2019,Oberg2019}, highlighting that giant-planet formation in the outer disc remains a viable possibility. Alternative pathways, including pebble accretion, planetesimal fragmentation, giant impacts, and planet formation in structured discs, have been proposed to accelerate growth, yet no single framework currently provides a fully satisfactory explanation for giant-planet formation across all orbital separations \citep[e.g.,][and references therein]{Mordasini2024,Burn2025}. Our model does not prescribe the specific mechanism responsible for core growth. Instead, we assume that a sufficiently massive core has formed and focus on how different growth and migration histories affect the resulting primordial atmospheric composition. A fully self-consistent treatment of the core-growth phase within the adopted chemical-evolution framework is left to future work.

\section{Conclusions}\label{sec:concl}
%Summary:In this study, we investigate how planet formation in evolving protoplanetary discs shapes the primordial composition of giant-planet atmospheres. Our goal is to establish a physically motivated, multi-element framework linking the observed chemical properties of planetary atmospheres to their formation and migration histories and to the time-dependent physical and chemical evolution of their natal disc. We combined $N$-body simulations of planet formation with one-dimensional models of disc evolution. The planet formation model tracks mass growth, migration, and gas-planetesimal accretion within the core-accretion paradigm, while the disc model follows the viscous evolution of gas, radial drift of dust grains of representative sizes (0.1, 20, and 100 \si{\micro\metre}), and time-dependent volatile chemistry under four chemical setups: inheritance and reset, each coupled with low and high ionisation conditions. We simulated giant planets growing and migrating from a range of initial orbital distances down to 0.4~au, tracing the accretion of carbon, oxygen, nitrogen, and sulphur. The resulting atmospheric compositions were analysed in terms of elemental ratios normalised to stellar values: C/O*, C/N*, N/O*, S/N*, and C/S*. When all tracers are measurable, comparative analysis of these ratios reveals three compositional fingerprints associated with distinct enrichment regimes:

We present a physically motivated, multi-element framework that links the primordial atmospheric composition of giant planets to their formation and migration histories in evolving protoplanetary discs. By coupling planet formation and migration models with time-dependent disc evolution, radial dust drift, and volatile chemistry, we trace how carbon, oxygen, nitrogen, and sulphur are incorporated into primordial planetary atmospheres under different chemical and transport regimes. We analyse the atmospheric composition of six giant planets, growing and migrating from a range of initial orbital distances down to 0.4~au, in terms of elemental ratios normalised to stellar values: C/O*, C/N*, N/O*, S/N*, and C/S*. The study reveals three compositional fingerprints associated with distinct accretion regimes:

\begin{itemize}
\item Gas-dominated (GD) atmospheres, with metallicity set by gas accreted outside drift-enriched regions, exhibit N/O*~>~C/O*~>~C/N*, with superstellar N/O*, near-stellar C/O*, and substellar C/N* ratios, with either undetected S-bearing species or a substellar S/N* (near-stellar C/S*) ratio.
\item Planetesimal-dominated (PD) atmospheres, characteristic of extensive migration and substantial solid accretion, exhibit N/O*~<~C/O*~<~C/N*, S/N*~$\geq$~C/N*, and C/S*~$\leq$~C/O*, with superstellar S/N*, stellar to superstellar C/N*, substellar to stellar C/O* and C/S*, and substellar N/O* ratios. 
\item Drift-enhanced (DE) atmospheres, shaped by the accretion of volatile-enriched gas in discs with efficient radial drift, exhibit a volatile signature similar to that of planetesimal-dominated atmospheres (N/O*~<~ C/O*~<~ C/N*), but with a significantly higher volatile-to-refractory ratio, reflected in strongly superstellar C/S* ratios, consistent with predictions from pebble-accretion models.
\item The analysis of the elemental abundances C/H*, O/H*, N/H*, and S/H* confirms the same atmospheric classes and compositional signatures identified from elemental ratios. While elemental abundances provide complementary information in some regimes, they do not fundamentally alter the interpretation based on elemental ratios.
\end{itemize}

The identified compositional fingerprints remain broadly robust across all initial chemical and ionisation conditions explored. We further find that:

\begin{itemize}
\item Gas-dominated atmospheres are most sensitive to the disc’s initial chemical state: formation in reset discs yields lower C/O* ratios than in inheritance discs. By contrast, in planets that also accrete solids, atmospheric composition is governed primarily by migration history rather than initial chemistry.
\item The N/O*, C/N*, and S/N* ratios generally increase with migration extent, although degeneracies arise for planets forming beyond the CO and N$_2$ snowlines, while the C/O* ratio remains largely insensitive to migration, consistent with our previous results based on stationary disc models.
\item In the absence of nitrogen constraints, markedly superstellar C/S* ratios indicate drift-enhanced gas accretion, while substellar C/S* and C/O* ratios suggest dominant solid accretion. Hybrid gas-solid accretion scenarios remain partially degenerate, and migration constraints are weakened.
\item Atmospheric compositional signatures traditionally associated with drift-driven volatile enrichment are not unique to scenarios involving mm- to cm-sized pebbles. Similar qualitative patterns in both elemental ratios and elemental abundances can arise from the inward drift of sub-mm-sized grains and the subsequent release of volatiles at snowlines, although the absolute enrichment levels remain sensitive to the properties and evolutionary state of the disc.
\item The traditional link between planetary metallicity and the mass of accreted solids breaks down under efficient drift-driven atmospheric enrichment. Volatile-to-refractory ratios provide the most robust diagnostics of the dominant enrichment pathway. However, while the C/S* ratio efficiently traces drift-driven ice sublimation, the N/S* ratio more directly constrains the contribution of planetesimals.
\end{itemize}

\noindent This work establishes a physically motivated framework linking the elemental compositions of giant-planet atmospheres to their accretion and migration histories, as well as to the properties of their formation environments. By identifying robust compositional signatures and predictive trends across a range of disc chemistries and transport regimes, it provides a physically grounded basis and testable predictions to guide the interpretation of current and future spectroscopic observations with JWST, Ariel, and other forthcoming facilities, such as ELT. 
% This degeneracy can be broken by the volatile-to-refractory ratio, which remains sensitive to whether heavy elements are delivered primarily through volatile-rich gas or solids \citep{Chachan2023,Crossfield2023}. 

\begin{acknowledgements}
    This work was carried with the support of the European Research Council via the Horizon 2020 Framework Programme ERC Synergy “ECOGAL” Project GA-855130 and the Italian Space Agency (ASI) through the ASI-INAF grants 2016-23-H.0 and 2021-5-HH.0. C.W.~acknowledges financial support from the Science and Technology Facilities Council and UK Research and Innovation (grant numbers ST/X001016/1 and MR/T040726/1). The computational resources were supplied by the Genesis cluster at INAF-IAPS and the technical support of Scigé John Liu is gratefully acknowledged. Finally, we thank the anonymous referee for their constructive comments and suggestions, which helped improve the clarity and quality of this manuscript.
\end{acknowledgements}

\bibliography{references.bib}
\bibliographystyle{aa}

\begin{appendix} 
\onecolumn
\section{Composition of the disc gas}\label{app:A}

\begin{figure*}
\centering
\includegraphics[width=0.95\textwidth]{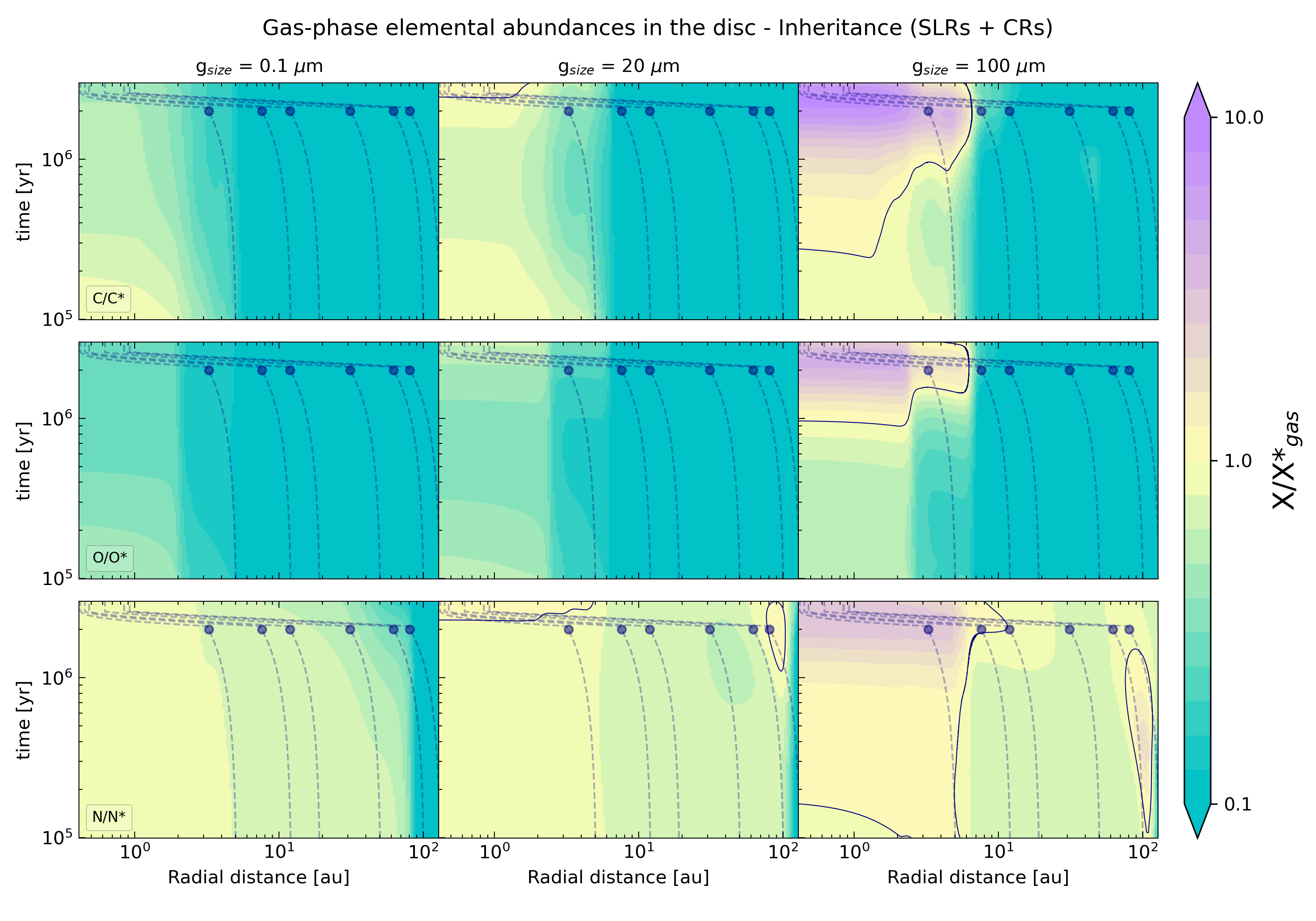}
\caption{Gas-phase elemental abundances of C (top row), O (middle row), and N (bottom row) in the natal disc as a function of radius and time. All abundances are normalised to their respective stellar values, with the dark solid line marking the stellar level ($=1$) in each panel. Results correspond to the low-ionisation inheritance scenario and to three dust grain sizes: 0.1~\si{\micro\meter} (left column), 20~\si{\micro\meter} (centre), and 100~\si{\micro\meter} (right). Dashed lines show the migration tracks of the six simulated planets, and the dark-blue dots indicate the onset of their runaway gas accretion phase. The comparison across columns illustrates the growing influence of radial drift, which produces increasingly strong volatile enrichments in the gas phase for larger grains.} 
\label{fig:disc_comp_ih}
\end{figure*}

\begin{figure*}
\centering
\includegraphics[width=0.95\textwidth]{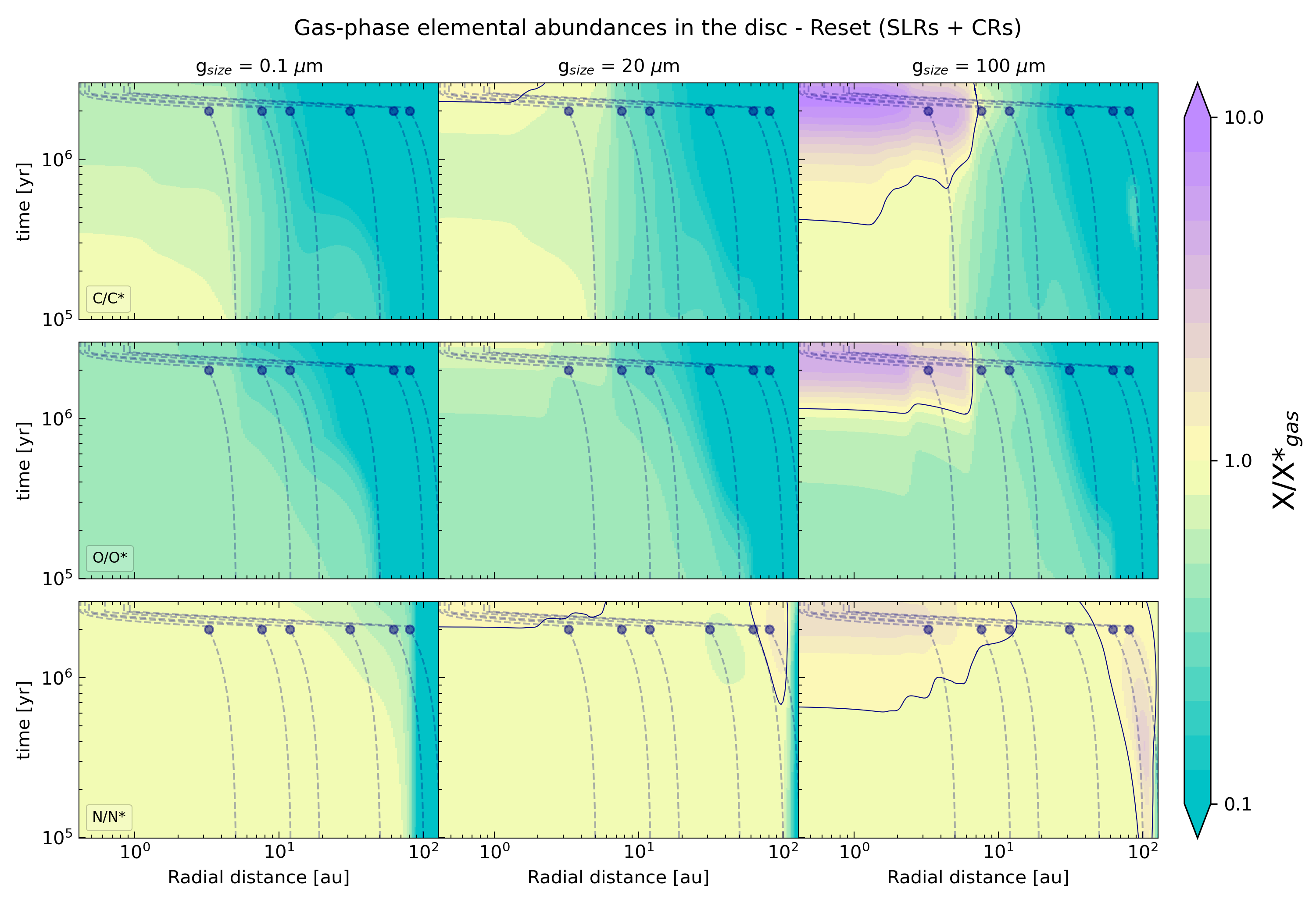}
\caption{Same as Fig.~\ref{fig:disc_comp_ih}, showing elemental abundances in the high-ionisation reset scenario.} 
\label{fig:disc_comp_rh}
\end{figure*}

Figures~\ref{fig:disc_comp_ih} and \ref{fig:disc_comp_rh} show the radial and temporal evolution of the gas-phase elemental abundances of C, O, and N in the protoplanetary disc, normalised to stellar values, as computed with the {\sc JADE} code \citep{Pacetti2025}. Each plot shows the evolution of the gas composition driven by the combined effects of chemical processing, gas transport, radial dust drift, and ice sublimation across snowlines, providing the disc-level physical foundation for the compositional fingerprints discussed throughout the paper. Results are shown from $10^5$~yr onward, after the onset of planetesimal formation, which converts 50$\%$ of the dust mass into a chemically inert reservoir. The panels correspond to three representative dust grain sizes (0.1, 20, and 100~\si{\micro\meter}) in the inheritance (Fig.~\ref{fig:disc_comp_ih}) and reset (Fig.~\ref{fig:disc_comp_rh}) chemical scenarios under high ionisation conditions. Dashed lines indicate the migration tracks of the six simulated planets, with dark-blue markers denoting the onset of runaway gas accretion.

The figures illustrate the background gas composition sampled by planets during their growth and migration pathways. In discs with efficient radial transport (100~\si{\micro\meter}-sized grains), the inward drift of icy solids and the subsequent release of volatiles at snowlines produce strong volatile enrichment of the inner disc on Myr timescales. In these scenarios, planets originating at smaller orbital distances ($R_0 \lesssim 20$~au, or $\lesssim 0.12,R_{\rm c}$ in our simulations) reach runaway gas accretion within regions already enriched by drift-driven ice sublimation, and therefore accrete high-metallicity gas.
By contrast, in discs with small grains (0.1~\si{\micro\meter}), the evolution is dominated by early inner-disc volatile depletion caused by planetesimal formation, which sequesters ices into an inert reservoir and reduces the volatile flux transported across snowlines (see \citealp{Pacetti2025} for further details).

\FloatBarrier 

\section{Planetary elemental ratios across different disc chemical scenarios}\label{app:B}

\begin{figure*}
\centering
\includegraphics[width=0.95\textwidth]{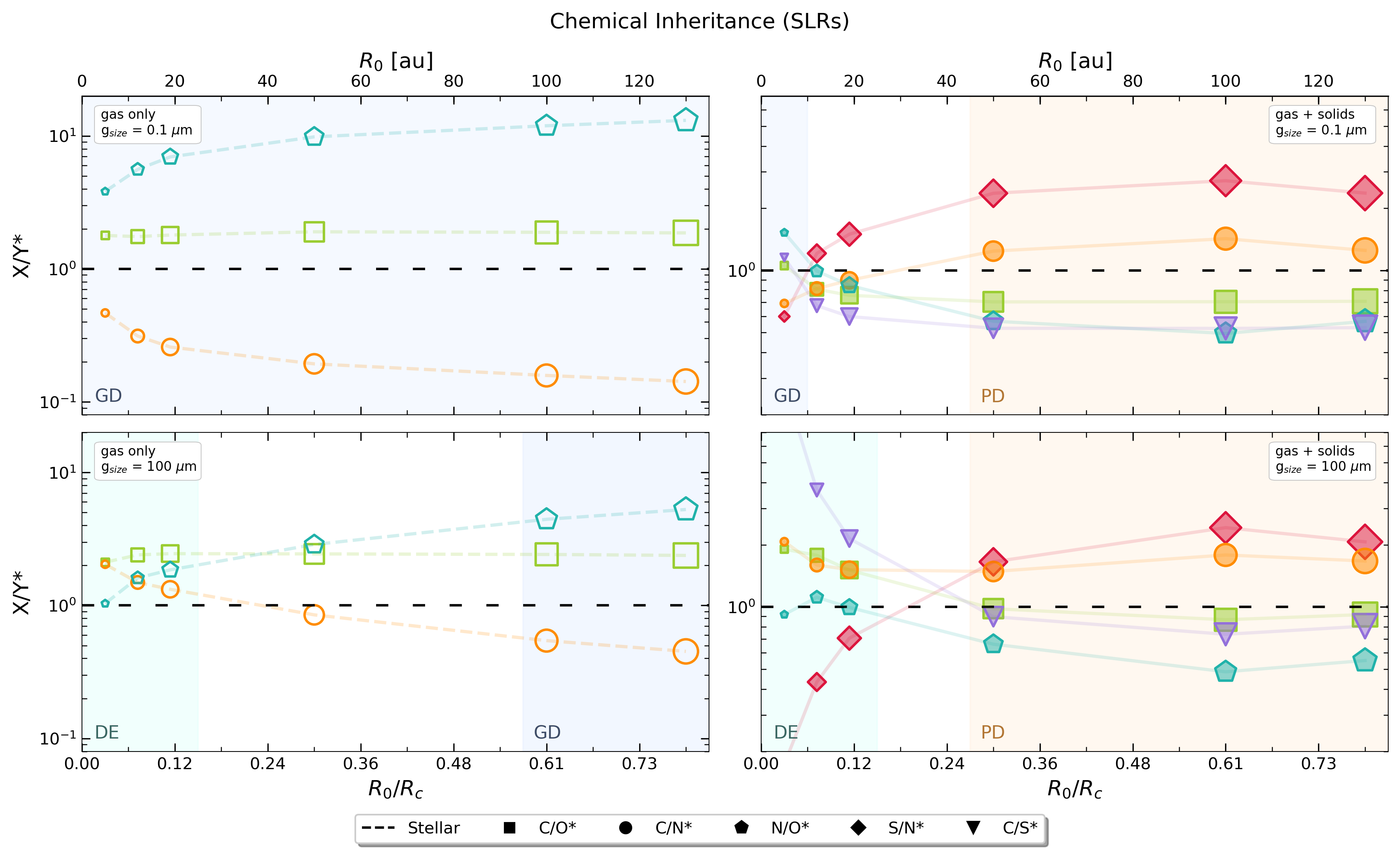}
\caption{Same as Fig.~\ref{fig:ternary}, showing planetary elemental ratios in the low-ionisation inheritance scenario.} 
\label{fig:ternary_il}
\end{figure*}

\begin{figure*}
\centering
\includegraphics[width=0.95\textwidth]{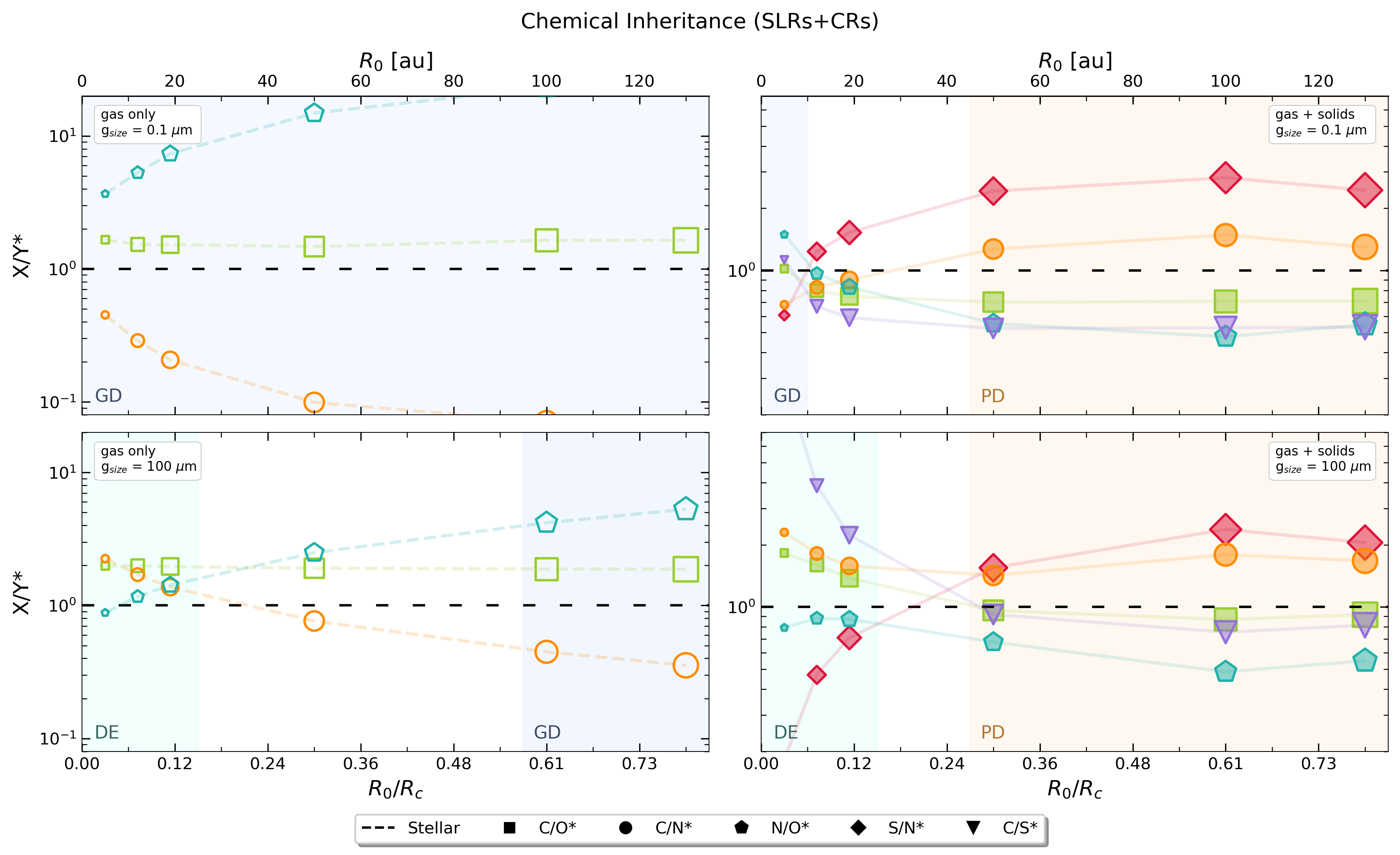}
\caption{Same as Fig.~\ref{fig:ternary}, showing planetary elemental ratios in the high-ionisation inheritance scenario.} 
\label{fig:ternary_ih}
\end{figure*}

\begin{figure*}
\centering
\includegraphics[width=0.95\textwidth]{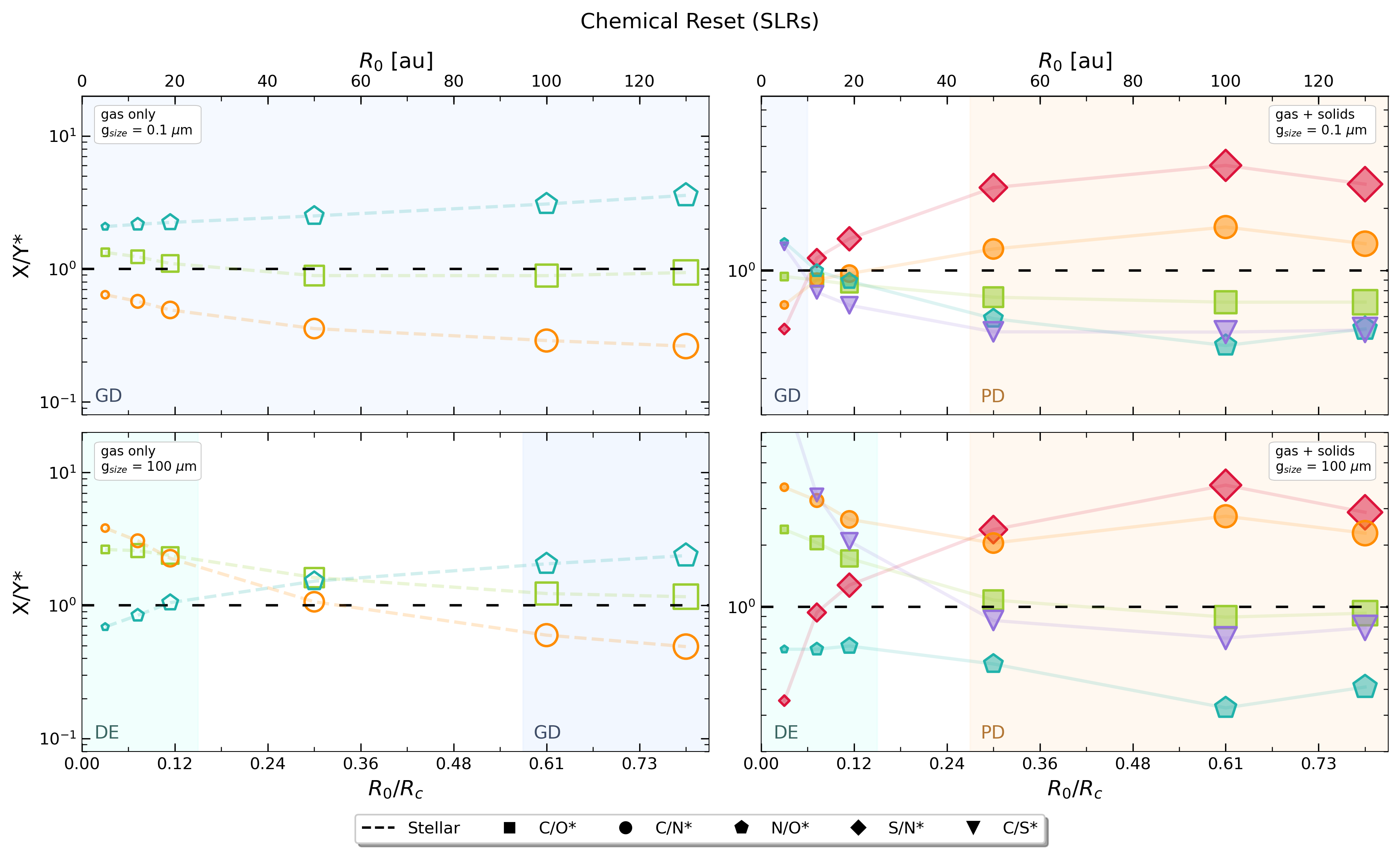}
\caption{Same as Fig.~\ref{fig:ternary}, showing planetary elemental ratios in the low-ionisation reset scenario.} 
\label{fig:ternary_rl}
\end{figure*}

Figures~\ref{fig:ternary_il}–\ref{fig:ternary_rl} present the same planetary elemental-ratio diagnostics shown in Fig.~\ref{fig:ternary}, extended to the other disc chemical scenarios explored in this work: low-ionisation inheritance, high-ionisation inheritance, and low-ionisation reset. Along with the high-ionisation reset case presented in the main text, these figures provide a complete overview of the predicted atmospheric compositions across all chemical setups considered in our simulations.

The three compositional classes identified in the main text – gas-dominated (GD), planetesimal-dominated (PD), and drift-enhanced (DE) – are consistently recovered across all disc chemical scenarios. While the absolute values of individual elemental ratios vary with ionisation state and chemical inheritance or reset conditions, the qualitative trends, the separation between enrichment regimes, and the overall diagnostic power of the combined elemental ratios are largely preserved. Differences between scenarios primarily affect the magnitude of atmospheric enrichment and the relative behaviour of nitrogen-bearing ratios, reflecting the varying partitioning of nitrogen between gas and ice phases and its transport during disc evolution. These effects introduce quantitative shifts and partial overlaps between classes in specific regimes but do not alter the underlying classification framework. 

These results demonstrate the robustness of the formation fingerprints identified in the main text and show that the inferred enrichment regimes are not specific to a single disc chemistry setup, but instead emerge naturally from the coupled processes of disc evolution, radial transport, migration, and accretion.

\FloatBarrier

\section{Elemental abundances in primordial atmospheres}\label{app:C}

\begin{figure*}
\centering
\includegraphics[width=0.95\textwidth]{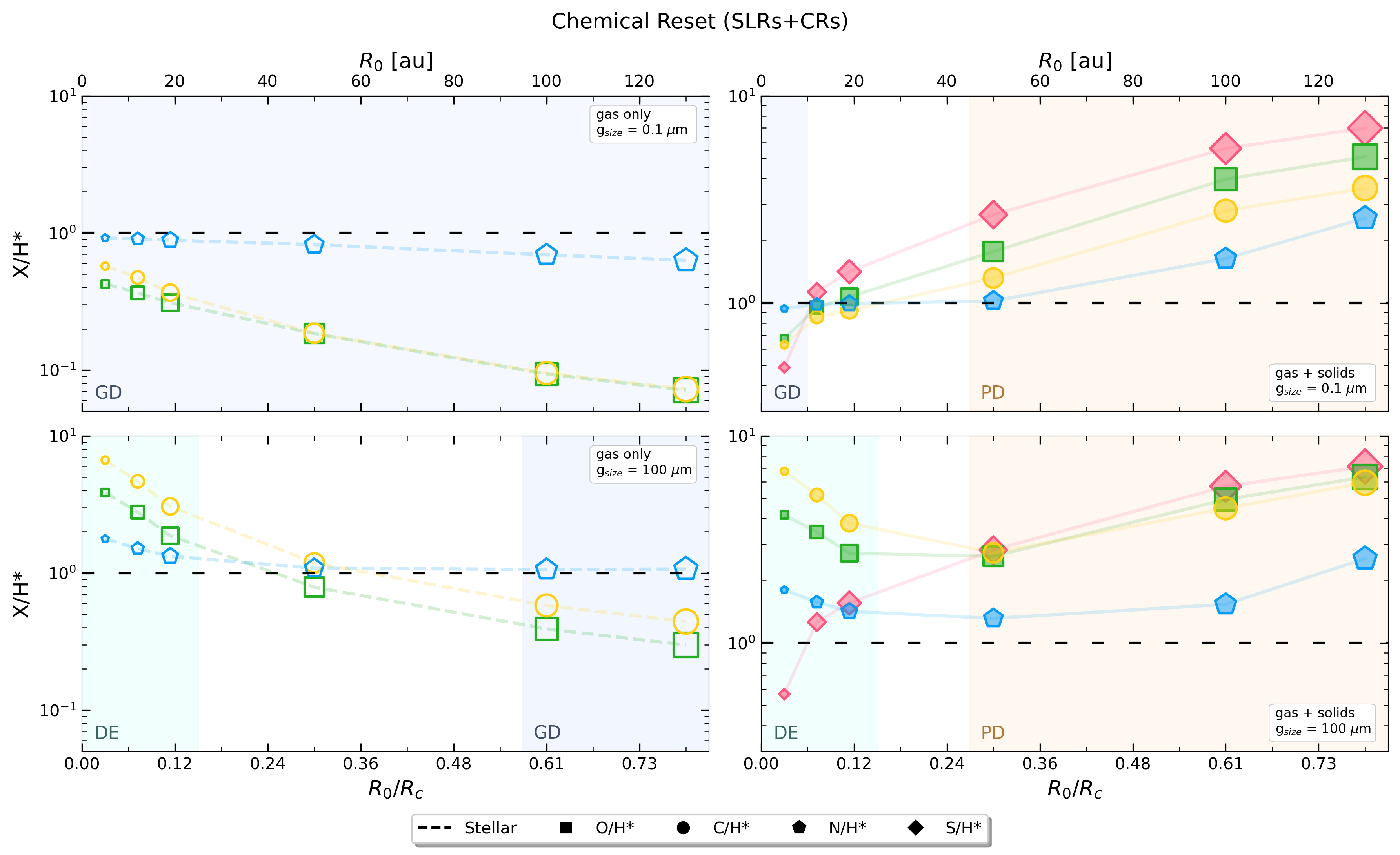}
\caption{Same as Fig.~\ref{fig:ternary}, showing elemental abundances in the primordial atmospheres of the six simulated giant planets as a function of their initial orbital positions. All abundances are normalised to their respective stellar value; the horizontal dashed line at 1 marks the stellar reference in all plots.} 
\label{fig:ternary_abs}
\end{figure*}

We extended the analysis presented in Sect.~\ref{sec:res_pf} to the elemental abundances C/H*, O/H*, N/H*, and S/H* in the primordial atmospheres of the six simulated giant planets, where the asterisk denotes normalisation to the stellar value. Figure~\ref{fig:ternary_abs} shows these abundances for the high-ionisation reset scenario, in a format analogous to Fig.~\ref{fig:ternary}. The overall picture is consistent with that derived in the main text from the elemental-ratio analysis. We identify the same three classes of primordial atmospheres: Gas-Dominated (GD), Planetesimal-Dominated (PD), and Drift-Enhanced (DE), while mixed atmospheres remain compositionally degenerate.

GD atmospheres (blue-shaded) are systematically characterised by substellar elemental abundances. We find C/H*$\sim$O/H*, both lower than N/H*, while S/H* is negligible in gas-only cases and remains substellar in GD atmospheres with a minor contribution from planetesimal accretion. The C/H* and O/H* abundances decrease with increasing migration distance, reflecting the progressively lower C and O content of the accreted gas. This trend is not apparent in the elemental-ratio framework, where the C/O* ratio remains nearly stellar across the entire GD population. However, the absolute enrichment levels are also sensitive to the degree of drift-driven volatile enrichment in the natal disc. GD atmospheres formed in discs with efficient radial drift can reach higher C/H* and O/H* values than their counterparts formed in weak-drift regimes, partially offsetting the effect of migration. Consequently, C/H* and O/H* do not provide an unambiguous constraint on migration extent within the GD class, although they retain information not captured by elemental ratios alone.

PD atmospheres (orange-shaded) are characterised by superstellar abundances of all elements, which increase with increasing migration distance due to the larger cumulative contribution of volatile and refractory material locked in planetesimals. In discs weakly affected by drift-driven ice sublimation (top-right panel), we find N/H*~<~C/H*~<~O/H*~<~S/H*, reflecting the increasing abundance of these elements in the accreted solids\footnote{Planetesimals carry the entire sulphur budget, more than half of the initial O and C budget – stored in refractory and semi-refractory material, as well as in ices, of which H$_2$O is the most abundant – while containing comparatively less N (see Sect.~\ref{sec:discmodel}).}. In the simulations with large grains (lower-right panel), the additional accretion of volatile-enriched gas modifies this trend by increasing O/H* and C/H* relative to the weak-drift regime, bringing them closer to S/H*, while N/H* remains comparatively lower due to the smaller fraction of N retained in ices and released into the gas. As a result, these atmospheres are characterised by S/H*~$\sim$~C/H*~$\sim$~O/H*, all significantly larger than N/H*. In scenarios where only C and O abundances can be observationally constrained, the superstellar C/H* and O/H* abundances may help distinguish PD atmospheres enriched by both planetesimals and volatile-enriched gas (lower-right panel) from GD atmospheres with only a minor contribution from planetesimals (top-right panel), since both classes can exhibit near-stellar C/O* ratios (see Sect.~\ref{sec:COS}). However, superstellar C/H* and O/H* are also characteristic of DE atmospheres. Consequently, the distinction between PD and DE atmospheres still requires information on the relative enrichment of volatile and refractory elements, as discussed below.

DE atmospheres (cyan-shaded) are characterised by superstellar O/H*, C/H*, and N/H* resulting from the accretion of volatile-enriched gas, with N/H*~<~O/H*~<~C/H*. The enrichment levels increase as planets form closer to the star and accrete material from regions increasingly affected by radial drift. In our simulations, DE and PD atmospheres can reach comparable levels of volatile enrichment in C, O, and N. The most robust distinction between the two classes arises from the relative enrichment of volatile and refractory tracers (see Sect.~\ref{sec:metallicity}). In particular, S/H* remains negligible or at most stellar to substellar in DE atmospheres because their enrichment is driven primarily by volatile-rich gas, whereas PD atmospheres reach strongly superstellar S/H* values through the accretion of refractory-rich planetesimals. The two classes remain separable as long as the planets do not accrete material from interior to the evaporation fronts of S-bearing refractory species. In that case, the set of considered tracers would need to be extended to more refractory elements sublimating interior to the planet's final orbit, which would better trace the accretion of solid material.

Mixed atmospheres remain degenerate also in the elemental-abundance framework. These atmospheres are generally characterised by near-stellar elemental abundances and do not display a unique compositional pattern. Comparison of elemental abundances therefore does not break the degeneracies already identified in the elemental-ratio analysis.

\FloatBarrier

\end{appendix}

\end{document}